\documentclass[lettersize,journal]{IEEEtran}
\usepackage{amsmath,amsfonts}
\usepackage{algorithm}
\usepackage{algorithmic}
\usepackage{array}
\usepackage[caption=false,font=normalsize,labelfont=sf,textfont=sf]{subfig}
\usepackage{graphicx}
\usepackage{hyperref}
\usepackage{multicol,multienum}
\usepackage{multirow}
\usepackage{textcomp}
\usepackage{stfloats}
\usepackage{url}
\usepackage{verbatim}

\def\BibTeX{{\rm B\kern-.05em{\sc i\kern-.025em b}\kern-.08em
		T\kern-.1667em\lower.7ex\hbox{E}\kern-.125emX}}
\usepackage{balance}
\begin{document}

\title{Column-Wise Analog Processing for Hybrid Precoding in Millimeter Wave Downlink Multi-User Massive MIMO Systems}
\author{Yu Zhang, Ziyang Meng, and Fangfang Yin
	\thanks{Y. Zhang is with the Department of Electronic Engineering, Tsinghua University, Beijing 100084, China (e-mail: zhangyu06@mail.tsinghua.edu.cn).}
	\thanks{Z. Meng is with the Key Laboratory of Space Utilization, Technology and Engineering Center for Space Utilization, Chinese Academy of Sciences, Beijing 100094, China, and with the University of Chinese Academy of Sciences, Beijing 100049, China (e-mail: mengziyang21s@ict.ac.cn).}
	\thanks{F. Yin is with the State Key Laboratory of Media Convergence and Communication, School of Information and Communication Engineering, Communication University of China, Beijing 100024, China (email: yinff@bupt.edu.cn).}
\vspace{-1em}
}

%\markboth{Journal of \LaTeX\ Class Files,~Vol.~18, No.~9, September~2020}%
%{How to Use the IEEEtran \LaTeX \ Templates}
	
\maketitle

\begin{abstract}

In millimeter wave (mmWave) massive MIMO systems, existing alternating minimization (AltMin) based hybrid digital-analog precoding algorithms can achieve near-optimal spectral efficiency (SE) of the fully-digital precoding. However, this kind of AltMin algorithms require several hundreds of iterations to optimize the analog precoder, thus increasing the complexity. This paper focuses on reducing complexity of the analog precoding and proposes a column-wise analog precoding (CWAP) algorithm. The main idea is to seek closed-form solution of the analog precoder, through which the analog precoder can be easily computed in one step instead of iterations, thus reducing the complexity. Specifically, by assuming that perfect digital precoder is deployed at the base station (BS) to eliminate interferences among users, we simplify the expression of achievable SE for each user. Subsequently, the simplified SE is further converted to the sum of a series of sub-rates, each of which is related to the corresponding column of the analog precoder. The optimization problem of maximizing the SE is then transformed into a series of sub-problems of maximizing each sub-rate. Upon solving each sub-problem, closed-form solution of each column of the analog precoder can be directly obtained without iterations, resulting in reduced complexity. After obtaining the analog precoder, our previously-proposed adaptive gradient algorithm is adopted to design the hybrid combiner for each user. Simulation results demonstrate that (a) when the number of RF chains equals the number of data streams, the proposed scheme can achieve approximately same sum-rate as the AltMin algorithms; (b) when the number of RF chains is larger than the number of data streams, the proposed scheme can achieve higher sum-rate than the AltMin algorithms; (c) the proposed scheme has lower complexity than the AltMin algorithms (almost one order of magnitude reduction in some cases).

\end{abstract}


\section{Introduction}

Due to sufficient spectrum from 30 GHz to 300 GHz, millimeter wave (mmWave) communication has been regarded as a primary candidate for future cellular networks, meeting the requirement of high-rate data transmission \cite{Hong2021TheRO, Li2022MobilitySF, Wang2023OnTR, Xue2024ASO}. The main obstacle for the successful deployment of mmWave systems is that the transmitted signals will undergo severe path loss due to ten-fold increase of carrier frequency \cite{Zhang2020Measurement5M, Xing2021MillimeterWA, Tang2022PathLM}. Thanks to the small wavelength of mmWave signals, massive multi-input multi-output (MIMO) can be deployed in a relatively small area to provide spatial diversity gain and combat the path loss. Meanwhile, MIMO precoding can be leveraged to support multi-stream data transmission for the single-user scenario and simultaneous communications between the base station (BS) and users for the multi-user scenario.

There are two main precoding architectures in current research literature, namely fully-digital and hybrid digital-analog precoding architectures \cite{Heath2016AnOO}. The traditional massive MIMO precoding is typically accomplished at baseband through a fully-digital precoder, which can adjust both the magnitude and phase of the generated signals. However, for the full-digital architecture, the needed radio frequency (RF) chains, including digital-to-analog converters (DACs), mixers and power amplifiers, are comparable in number to the antenna elements. Because the number of antennas as well as RF chains is usually very large, the fully-digital structure has been turned out to be unsuitable for mmWave massive MIMO systems due to prohibitive power consumption of RF chains \cite{Heath2016AnOO}.

At the mmWave band, due to few scatterers, the number of propagation paths is usually very small, which leads to low-rank property of the channel matrix and sparsity of the mmWave channel \cite{Rangan2014MillimeterCW}. The rank of the channel matrix determines the upper bound of the number of RF chains. Thus a hybrid digital-analog precoding architecture is proposed in \cite{Ayach2014SpatiallySP}, which only requires a small number of RF chains interfacing between a low-dimensional digital precoder and a high-dimensional analog precoder. Due to reduced number of RF chains, power consumption and hardware cost of the hybrid structure are much lower than those of the fully-digital structure. Therefore, hybrid structure is more suitable for mmWave systems.

For the hybrid structure, the analog precoder is usually composed of a network of constant-magnitude analog phase shifters (APSs). When the transmitted signal is processed by the analog precoder, only the phase can be adjusted and the amplitude remains constant. In other words, the adjustable range of the transmitted signal by analog precoder is constrained. Compared with fully-digital architecture, hybrid architecture has caused a loss of the spectral efficiency (SE) to some extent due to the constant-magnitude constraint of the analog precoder. In the current literature, the objective of hybrid precoding is to minimize the Euclidean distance between the hybrid precoder and the fully-digital precoder with acceptable complexity \cite{Liu2022ADC, Luo2022MDLAH}.

The existing hybrid precoding algorithms can be divided into two categories, i.e., codebook-based algorithms and non-codebook-based algorithms. As for codebook-based hybrid precoding, the problem of hybrid precoding is viewed as a sparse signal recovery problem \cite{Ayach2014SpatiallySP}. An algorithmic solution is given in \cite{Ayach2014SpatiallySP} by using orthogonal matching pursuit (OMP), where each column of the analog precoder is selected from a pre-defined codebook. In order to reduce complexity of the OMP algorithm, the variants of the OMP algorithm are proposed, including the parallel-index-selection matrix-inversion-bypass simultaneous orthogonal matching pursuit (PIS-MIB-SOMP) algorithm \cite{Lee2015AHR}, the order-recursive least squares generalized orthogonal matching pursuit (ORLS-gOMP) algorithm \cite{Zhang2018LowCH} and the orthogonality based matching pursuit (OBMP) algorithm \cite{Hung2015LowHP}. Because the pre-defined codebooks in \cite{Ayach2014SpatiallySP, Lee2015AHR, Zhang2018LowCH, Hung2015LowHP} constraint the value of each entry of the analog precoding matrix, the optimal solution of the analog precoder is not reached. There still exists a large gap between the achievable SE of the codebook-based hybrid precoding and fully-digital precoding.

In order to approach the SE of the fully-digital precoding, non-codebook-based algorithms are put forward by alternately optimizing the digital- and analog- precoders to minimize the Euclidean distance between the hybrid precoder and fully-digital precoder, namely alternating minimization (AltMin) algorithms. The hybrid precoding is regarded as a matrix decomposition problem and optimization algorithms are adopted to design the hybrid precoder. The digital precoder is first updated according to the well-known least square solution in the external iteration. Then by fixing the digital precoder, the analog precoder can be optimized via conjugate gradient (CG) \cite{Yu2016AlternatingMA}, phase extraction (PE) \cite{Yu2016AlternatingMA}, interior-point method \cite{Ni2017Near-optimalHP}, Barzilai-Borwein (BB) gradient \cite{Mulla2020Barzilai-BorweinGA} or gradient projection (GP) \cite{Chen2019GradientPA} in the internal iteration. The digital- and analog- precoders are alternately optimized until the stop condition is satisfied, i.e., the Euclidean distance between the hybrid precoder and fully-digital precoder is smaller than a sufficiently small positive value, or the number of iterations exceeds the pre-defined threshold. Although achieving near-optimal performance of the fully-digital precoding, the AltMin algorithms usually require several hundreds of iterations to optimize the analog precoder and thus have high complexity.

Most of the aforementioned literature generally assume that the resolution of APSs are infinite. However, in practical applications, APSs are controlled digitally and the phase resolution depends on the quantized bit \cite{Lin2017OnTQ}. The power consumption of APSs increases with the increase of resolution. Constrained by the power consumption and hardware cost of APSs, low-resolution APSs are usually adopted in hybrid precoding structure \cite{Nouri2023HybridPB}. Based on finite-resolution APSs, Gram-Schmidt orthogonalization based greedy algorithm \cite{Alkhateeb2016FrequencySH}, heuristic hybrid beamforming design \cite{Sohrabi2016HybridDA}, iterative phase matching (IPM) algorithm \cite{Wang2018HybridPA}, coordinate descent method \cite{Chen2017HybridBW} and adaptive cross-entropy algorithm \cite{Zhang2021MachineLB, Zhang2022TreeAA} have been presented. However, the algorithms in \cite{Alkhateeb2016FrequencySH, Sohrabi2016HybridDA, Wang2018HybridPA} have resulted in reduced spectral efficiency and the algorithms in \cite{Chen2017HybridBW, Zhang2021MachineLB, Zhang2022TreeAA} have relatively-high complexity. Furthermore, the AltMin algorithms in \cite{Yu2016AlternatingMA, Ni2017Near-optimalHP, Mulla2020Barzilai-BorweinGA, Chen2019GradientPA} can also be applicable to the mmWave MIMO system with low-resolution APSs by converting the phase of each entry of the analog precoder and combiner into a quantized value. Though achieving high performance in terms of the SE, there is still room for reducing the complexity of the AltMin algorithms.

The motivation of this paper is to reduce complexity of the existing near-optimal AltMin hybrid precoding algorithms. Aiming at the mmWave downlink multi-user massive MIMO system with low-resolution APSs where both the BS and the users are equipped with multiple antennas and multiple data streams, the column-wise analog precoding (CWAP) algorithm is proposed. The analog precoder can be directly computed in one step through closed-form solution instead of iterations, thus reducing the complexity. The main contributions are summarized as follows:

\begin{itemize}
  \item We simplify the expression of the achievable SE based on the assumption that perfect digital precoder is deployed at the BS to cancel interferences among users. The simplified SE mainly depends on the analog precoder. Then the simplified SE is converted to the sum of a series of sub-rates, each of which is related to the corresponding column of the analog precoder. The original SE maximization problem is transformed into a series of sub-problems to maximize each sub-rate, which can be solved successively.
  \item Upon solving each sub-problem, we find the closed-form expression of each column of the analog precoder. Therefore, the analog precoder can be directly computed by the closed-form solution in one step without iterations. Due to avoiding iterative operations, the complexity of calculating the analog precoder can be reduced significantly.
  \item We apply our previously-proposed adaptive gradient (AG) algorithm in \cite{Zhang2021FrequencySH} to design the hybrid combiner for each user. The number of iterations is smaller than those of the AltMin algorithms. As a result, the complexity of designing the hybrid combiners can be reduced.
  \item Detailed simulations are executed to evaluate the effectiveness of the proposed scheme. Simulation results demonstrate that, when the number of RF chains is equal to or larger than the number of data streams, the sum-rate of the proposed scheme is approximately same as or higher than that of the AltMin algorithms with lower complexity (almost one order of magnitude reduction of the number of complex multiplications and divisions in some cases).
\end{itemize}

The rest of the paper is organized as follows. Section II describes the model of the considered mmWave downlink multi-user massive MIMO system. The proposed scheme as well as the complexity analysis are given in Section III. Section IV shows the simulation results and Section V concludes the paper.

\begin{figure*}[bt!]
	\centering
	\includegraphics[width=6in]{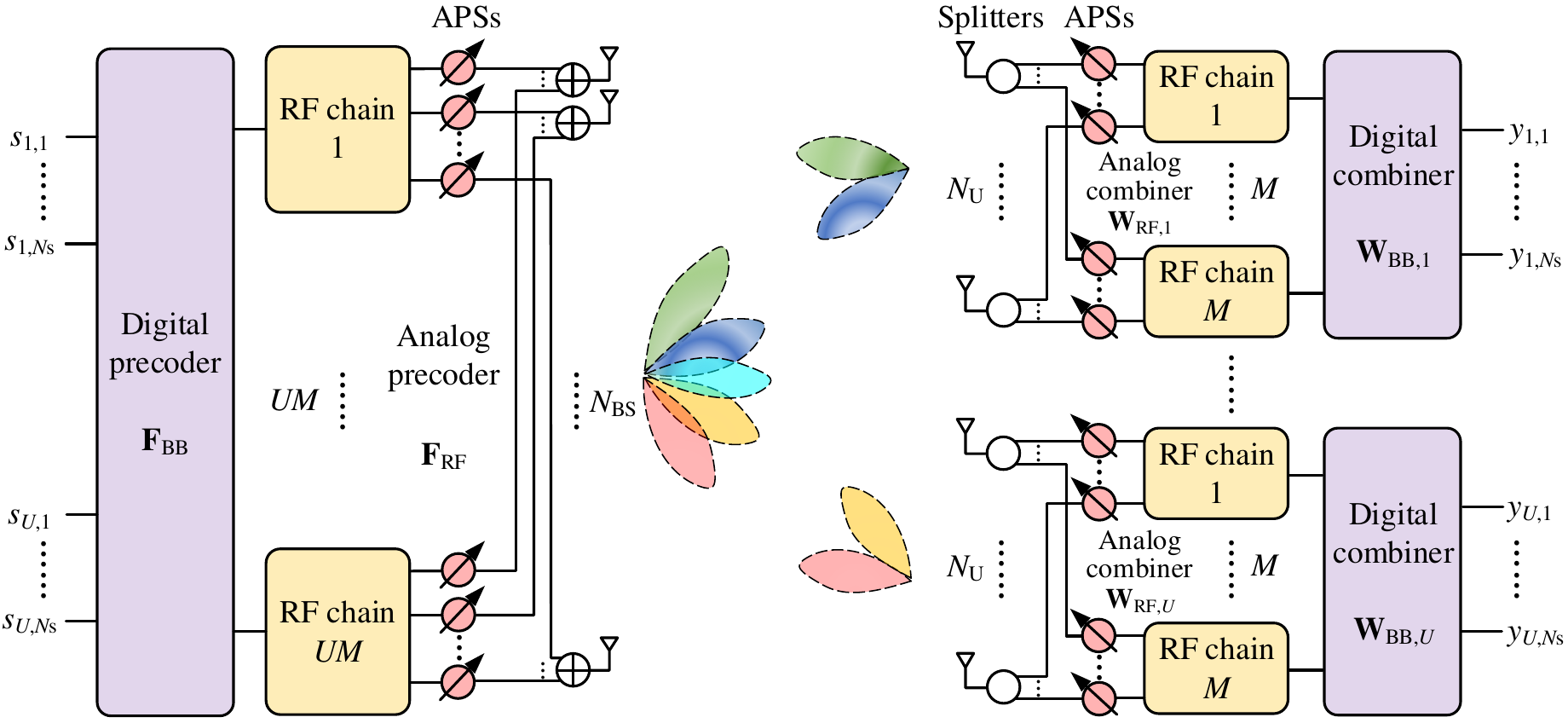}\\
	\caption{Block diagram of the considered mmWave downlink multi-user MIMO system.}
	\label{fig:fig-multiple-system-model}
\end{figure*}

\emph{Notations}: ${a}$ is a scalar, ${\mathbf{a}}$ is a vector and ${\mathbf{A}}$ is a matrix. ${{\mathbf{0}}_{M \times N}}$ is an ${M \times N}$ zero matrix and ${{\mathbf{I}}_N}$ is an ${N \times N}$ identity matrix. ${\mathbb{C}^{M \times N}}$ means an ${M \times N}$ complex matrix. $a^*$ is the conjugate of $a$. $\left| {\mathbf{A}} \right|$, ${{\mathbf{A}}^T}$, ${{\mathbf{A}}^H}$, ${{\mathbf{A}}^{-1}}$ and ${\left\| {\mathbf{A}} \right\|_F}$ represent the determinant, transpose, conjugate transpose, inversion and Frobenius norm of $\mathbf{A}$, respectively. $\mathbf{A} \otimes \mathbf{B}$ means the Kronecker product of $\mathbf{A}$ and $\mathbf{B}$. ${\mathbf{A}}\left( {m,n} \right)$ is the ${m}$-th row and ${n}$-th column entry of ${\mathbf{A}}$. $\mathcal{CN}\left( {{\mathbf{a}},{\mathbf{A}}} \right)$ denotes a complex Gaussian vector with mean ${\mathbf{a}}$ and covariance ${\mathbf{A}}$. $\mathbb{E}\left\{ {\mathrm{\cdot}} \right\}$ means the expectation.

\section{System Model}

As illustrated in Fig. \ref{fig:fig-multiple-system-model}, a multi-user downlink mmWave massive MIMO system is considered, where a BS communicates with $U$ users simultaneously. Each user is equipped with $N_{\rm{U}}$ antennas, $M$ RF chains and $N_{\rm{S}}$ data streams. The BS is equipped with $N_{\rm{BS}}$ antennas, $UM$ RF chains and $UN_{\rm{S}}$ data streams. At the BS, the $UN_{\rm{S}} \times 1$ transmitted signal ${\bf{s}} = [{\bf{s}}_1^T,...,{\bf{s}}_u^T,...,{\bf{s}}_U^T]^T$ is precoded via a $UM \times UN_{\rm{S}}$ digital precoder ${\bf{F}}_{\rm{BB}} = \left[ {{{\bf{F}}_{{\rm{BB}},1}}, \ldots , {{\bf{F}}_{{\rm{BB}},u}}, \ldots, {{\bf{F}}_{{\rm{BB}},U}}} \right]$ and an $N_{\rm{BS}} \times UM$ analog precoder ${\bf{F}}_{\rm{RF}} = \left[ {{{\bf{F}}_{{\rm{RF}},1}}, \ldots , {{\bf{F}}_{{\rm{RF}},u}}, \ldots, {{\bf{F}}_{{\rm{RF}},U}}} \right]$, and then mapped into the transmit antennas. ${\bf{s}}_u = [s_{u,1},...,s_{u,N_{\rm{S}}}]^T \in {\mathbb{C}}^{N_{\rm{S}} \times 1}$, ${{\bf{F}}_{{\rm{BB}},u}} \in {\mathbb{C}}^{UM \times N_{\rm{S}}}$ and ${{\bf{F}}_{{\rm{RF}},u}} \in {\mathbb{C}}^{N_{\rm{BS}} \times M}$ denote the transmitted signal, digital precoder and analog precoder for the $u$-th user. ${{\bf{F}}_{{\rm{RF}},u}}$ is constrained by constant-modulus APSs such that $\left| {{{\bf{F}}_{{\rm{RF}},u}}\left( {n,m} \right)} \right| = 1/\sqrt {{N_{{\rm{BS}}}}}$. ${{\bf{F}}_{{\rm{RF}}}}$ and ${{\bf{F}}_{{\rm{BB}}}}$ satisfy the power constraint such that $\left\| {{{\bf{F}}_{{\rm{RF}}}}{{\bf{F}}_{{\rm{BB}}}}} \right\|_F^2 = U{N_{\rm{S}}}$.

\begin{figure}[bt!]
	\centering
	\includegraphics[width=2in]{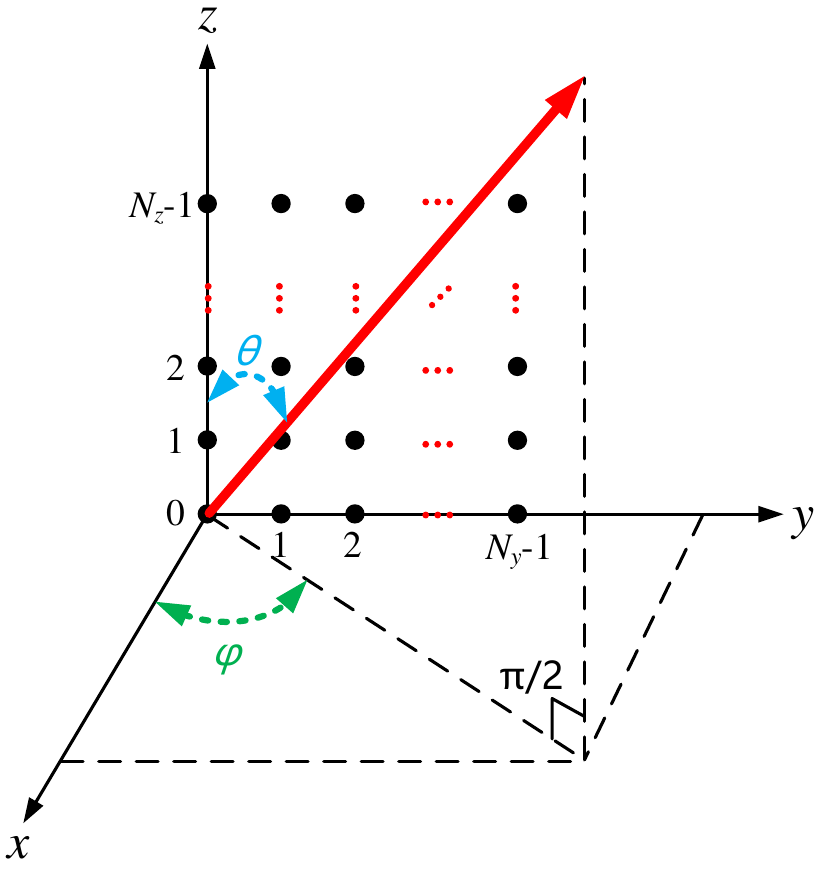}\\
	\caption{Illustration of UPA.}
	\label{fig:fig-UPA}
\end{figure}

At the $u$-th user, the received signal is decoded by an $N_{\rm{U}} \times M$ analog combiner ${\bf{W}}_{{\rm{RF}},u}$ and an $M \times N_{\rm{S}}$ digital combiner ${\bf{W}}_{{\rm{BB}},u}$. ${\bf{W}}_{{\rm{RF}},u}$ also satisfies the constant-modulus constraint such that $\left| {{{\bf{W}}_{{\rm{RF}},u}}\left( {n',m'} \right)} \right| = 1/\sqrt {{N_{\rm{U}}}}$. The combined signal can be written as
\begin{equation}\label{}
  	\begin{array}{l}
		{{\bf{y}}_u} = {\bf{W}}_{{\rm{BB}},u}^H{\bf{W}}_{{\rm{RF}},u}^H{{\bf{H}}_u}{{\bf{F}}_{{\rm{RF}}}}{{\bf{F}}_{{\rm{BB}},u}}{{\bf{s}}_u} + \\
		\underbrace {\sum\limits_{j = 1,j \ne u}^U {{\bf{W}}_{{\rm{BB}},u}^H{\bf{W}}_{{\rm{RF}},u}^H{{\bf{H}}_u}{{\bf{F}}_{{\rm{RF}}}}{{\bf{F}}_{{\rm{BB}},j}}{{\bf{s}}_j}} }_{{\rm{interference}}} + \underbrace {{\bf{W}}_{{\rm{BB}},u}^H{\bf{W}}_{{\rm{RF}},u}^H{{\bf{n}}_u}}_{{\rm{noise}}}
	\end{array} ,
\tag{1}
\end{equation}
where ${{\bf{H}}_u}$ is the ${N_{{\rm{U}}}} \times {N_{{\rm{BS}}}}$ channel matrix between the BS and the $u$-th user. ${{\bf{n}}_u} \sim \mathcal{CN}\left( {{\bf{0}},{\sigma _u ^2}{{\bf{I}}_{{N_{{\rm{U}}}}}}} \right)$ is the ${N_{{\rm{U}}}} \times 1$ Gaussian noise vector.
Due to limited scattering, ${{\bf{H}}_u}$ is depicted as a sum of $N_{\rm{p}}$ propagation paths and can be expressed as \cite{Saleh1987ASM}
\begin{equation}\label{equ:channel}
  	{{\bf{H}}_u} = \sqrt {\frac{{{N_{{\rm{BS}}}}{N_{\rm{U}}}}}{{{N_{\rm{p}}}}}} \sum\limits_{l = 1}^{{N_{\rm{p}}}} {{\alpha _l}{{\bf{a}}_{u,{\rm{R}}}}\left( {\varphi _{u,{\rm{R}}}^l,\theta _{u,{\rm{R}}}^l} \right){{\bf{a}}_{u,{\rm{T}}}}{{\left( {\varphi _{u,{\rm{T}}}^l,\theta _{u,{\rm{T}}}^l} \right)}^H}} ,
\tag{2}
\end{equation}
where ${\alpha}_l$ is the complex gain, ${\phi}_{u,{\rm{T}}}^l$ (${\theta}_{u,{\rm{T}}}^l$) is the azimuth (elevation) angle of departure (AoD), and ${\phi}_{u,{\rm{R}}}^l$ (${\theta}_{u,{\rm{R}}}^l$) is the azimuth (elevation) angle of arrival (AoA) of the $l$-th path, respectively. ${\bf{a}}_{u,{\rm{T}}}$ and ${\bf{a}}_{u,{\rm{R}}}$ are the array response vectors corresponding to the angles of departure and arrival. With regard to a uniform planar array (UPA) as shown in Fig. \ref{fig:fig-UPA}, the antenna elements are located in the $yoz$ plane. The spacing between each two antenna element is half a wavelength. $N_y$ and $N_z$ are the numbers of antenna elements along the $y$-axis and $z$-axis, respectively. The azimuth angle $\varphi$ denotes the angle between the projection of beam direction on the $xoy$ plane and $x$-axis. The elevation angle $\theta$ denotes the angle between the beam direction and $z$-axis. The expression of the array response vector of UPA is \cite{Ayach2014SpatiallySP, Yu2016AlternatingMA}
\begin{equation}\label{equ:array-response-vector}
  	\begin{array}{*{20}{c}}
		{{{\bf{a}}_{{\rm{UPA}}}}\left( {\varphi ,\theta } \right) = \frac{1}{{\sqrt {{N_y}{N_z}} }}\left[ {1,...,{e^{j\pi \left( {{n_y}\sin \varphi \sin \theta  + {n_z}\cos \theta } \right)}},} \right.}\\
		{{{\left. {...,{e^{j\pi \left( {({N_y} - 1)\sin \varphi \sin \theta  + ({N_z} - 1)\cos \theta } \right)}}} \right]}^T}} ,
	\end{array}
\tag{3}
\end{equation}
where ${n_y} = 0,1, \ldots ,{N_y-1}$ and ${n_z} = 0,1, \ldots ,{N_z-1}$ are the $y$- and $z$- axis indices of the antenna element.

Assuming that $\bf{s}$ follows a Gaussian distribution and satisfies $\mathbb{E}\left\{ \mathbf{s}{{\mathbf{s}}^{H}} \right\}=\left( \rho /U{{N}_{\text{S}}} \right){{\mathbf{I}}_{U{{N}_{\text{S}}}}}$, the achievable SE of the $u$-th user can be computed by \cite{Goldsmith2003CapacityLO}
\begin{equation}\label{equ:initial-Ru}
	\begin{array}{*{20}{c}}
		{{R_u} = {{\log }_2}\left| {{{\bf{I}}_{{N_{\rm{S}}}}} + \frac{\rho }{{U{N_{\rm{S}}}}}{\bf{R}}_u^{ - 1}{\bf{W}}_{{\rm{BB}},u}^H{\bf{W}}_{{\rm{RF}},u}^H{{\bf{H}}_u} } \right.}\\
		{\left. \times {{{\bf{F}}_{{\rm{RF}}}}{{\bf{F}}_{{\rm{BB}},u}}{\bf{F}}_{{\rm{BB}},u}^H{\bf{F}}_{{\rm{RF}}}^H{\bf{H}}_u^H{{\bf{W}}_{{\rm{RF}},u}}{{\bf{W}}_{{\rm{BB}},u}}} \right|} ,
	\end{array}
\tag{4}
\end{equation}
where $\rho$ is the average transmit power, and
\begin{equation}\label{equ:R_u}
  	\begin{array}{*{20}{c}}
		{{{\bf{R}}_u} = \frac{\rho }{{U{N_{\rm{S}}}}}\sum\limits_{j = 1,j \ne u}^U { 	{{\bf{W}}_{{\rm{BB}},u}^H{\bf{W}}_{{\rm{RF}},u}^H{{\bf{H}}_u}{{\bf{F}}_{{\rm{RF}}}}{{\bf{F}}_{{\rm{BB}},j}}{\bf{F}}_{{\rm{BB}},j}^H} } }\\
		{ \times {\bf{H}}_{{\rm{eq}},u}^H{{\bf{W}}_{{\rm{BB}},u}} +  {\sigma _u^2{\bf{W}}_{{\rm{BB}},u}^H{\bf{W}}_{{\rm{RF}},u}^H{{\bf{W}}_{{\rm{RF}},u}}{{\bf{W}}_{{\rm{BB}},u}}} .}
	\end{array}
\tag{5}
\end{equation}

The objective of multi-user hybrid precoding is to design the hybrid precoder at the BS and the hybrid combiners at the users to maximize the sum-rate of all the users subject to the power constraint and the constant-modulus constraint
\begin{equation}\label{equ:initial-objective}
\begin{array}{*{20}{c}}
	\begin{array}{l}
		\left( {\left\{ {{\bf{F}}_{{\rm{RF}},u}^{{\rm{opt}}},{\bf{F}}_{{\rm{BB}},u}^{{\rm{opt}}},{\bf{W}}_{{\rm{RF}},u}^{{\rm{opt}}},{\bf{W}}_{{\rm{BB}},u}^{{\rm{opt}}}} \right\}_{u = 1}^U} \right)\\
		 = \mathop {\arg \max }\limits_{\left\{ {{{\bf{F}}_{{\rm{RF}},u}},{{\bf{F}}_{{\rm{BB}},u}},{{\bf{W}}_{{\rm{RF}},u}},{{\bf{W}}_{{\rm{BB}},u}}} \right\}_{u = 1}^U} \sum\limits_{u = 1}^U {{R_u}}
		\end{array}\\
		{s.t.\;\;\left\| {{{\bf{F}}_{{\rm{RF}}}}{{\bf{F}}_{{\rm{BB}}}}} \right\|_F^2 = U{N_{\rm{S}}},}\\
		{\;\;\;\;\;\;\;\;\;\;\left| {{{\bf{F}}_{{\rm{RF}},u}}\left( {n,m} \right)} \right| = 1/\sqrt {{N_{{\rm{BS}}}}} ,}\\
		{\;\;\;\;\;\;\;\;\;\;\left| {{{\bf{W}}_{{\rm{RF}},u}}\left( {n',m'} \right)} \right| = 1/\sqrt {{N_{\rm{U}}}} .}
	\end{array}
\tag{6}
\end{equation}

\section{Proposed Hybrid Precoding Scheme}

Due to the constant-modulus constraints of ${\bf{F}}_{\rm{RF}}$ and ${\bf{W}}_{{\rm{RF}},u}$ imposed by the constant-amplitude APSs, the optimization problem \eqref{equ:initial-objective} is non-convex and intractable. To deal with the problem, hybrid precoder and combiners of all the users are designed separately. Firstly, by assuming that perfect digital precoder is deployed at the BS to cancel interferences among users, $R_u$ can be simplified to a function mainly depending on ${\bf{F}}_{{\rm{RF}},u}$. Then, $R_u$ is converted to the sum of a series of sub-rates, each of which is related to the corresponding column of ${\bf{F}}_{{\rm{RF}},u}$. The original SE optimization problem is transformed into a series of sub-problem of maximizing each sub-rate. Upon solving each sub-problem, the closed-form solution of each column of ${\bf{F}}_{{\rm{RF}},u}$ is found. Next, the equivalent baseband channel matrices are computed and the BD method \cite{Ni2016HybridBD} is adopted to obtain ${\bf{F}}_{\rm{BB}}$ and fully-digital combiner for each user ${{\mathbf{W}}_{u,\text{opt}}}$. Finally, our previously-proposed AG algorithm is applied to acquire ${\bf{F}}_{{\rm{RF}},u}$ and ${\bf{W}}_{{\rm{BB}},u}$ by minimizing the Euclidean distance between ${{\mathbf{W}}_{\text{RF},u}}{{\mathbf{W}}_{\text{BB},u}}$ and ${{\mathbf{W}}_{u,\text{opt}}}$.

\subsection{BD Based Digital Precoding}

Assume that each user adopts the fully-digital combiner, the matrix product ${{\mathbf{W}}_{\text{RF},u}}{{\mathbf{W}}_{\text{BB},u}}$ in \eqref{equ:initial-Ru} and \eqref{equ:R_u} can be replaced by ${{\mathbf{W}}_{u}} \in {\mathbb{C}}^{N_{\rm{U}} \times N_{\rm{S}}}$. Define the equivalent baseband channel of the $u$-th user as ${{\bf{H}}_{{\rm{eq}},u}} = {{\bf{H}}_u}{{\bf{F}}_{{\rm{RF}}}} \in {\mathbb{C}}^{N_{\rm{U}} \times UM}$. The BD method in \cite{Ni2016HybridBD} is used to obtain the digital precoder ${{{\bf{F}}_{{\rm{BB}},u}}}$ and the fully-digtal combiner ${{\mathbf{W}}_{u}}$ for each user.

We introduce a variable ${{\bf{\bar H}}_u}$ as
\begin{equation}\label{equ:Hu}
	{{\bf{\bar H}}_u} = {\left[ {{\bf{H}}_{{\rm{eq}},1}^T, \ldots ,{\bf{H}}_{{\rm{eq}},u - 1}^T,{\bf{H}}_{{\rm{eq}},u + 1}^T, \ldots ,{\bf{H}}_{{\rm{eq}},U}^T} \right]^T} .
	\tag{7}
\end{equation}
${{\bf{\bar H}}_u}$ is composed of all the users' equivalent baseband channels except for ${{\bf{H}}_{{\rm{eq}},u}}$. ${{{\bf{F}}_{{\rm{BB}},u}}}$ should be lied in the null space of ${{\bf{\bar H}}_u}$ to eliminate the interferences between users such that
\begin{equation}\label{equ:null}
	{{\bf{H}}_{{\rm{eq}},u}}{{\bf{F}}_{{\rm{BB}},j}} = {\bf{0}},\forall j \ne u .
	\tag{8}
\end{equation}

The singular value decomposition (SVD) of ${{\bf{\bar H}}_u}$ is
\begin{equation}\label{equ:SVDHu}
	{{{\bf{\bar H}}}_u} = {{{\bf{\bar U}}}_u}{{{\bf{\bar \Sigma }}}_u}{\left[ {{\bf{\bar V}}_u^{\left( {\left( {U - 1} \right)M} \right)},{\bf{\bar V}}_u^{\left( M \right)}} \right]^H} ,
	\tag{9}
\end{equation}
where ${{\bf{\bar V}}_u^{\left( {\left( {U - 1} \right)M} \right)}}$ and ${{\bf{\bar V}}_u^{\left( M \right)}}$ are the first $(U-1)M$ and rest $M$ right singular vectors of ${{\bf{\bar H}}_u}$. The we have
\begin{equation}\label{}
	{{\bf{H}}_{{\rm{eq}},u}}{\bf{\bar V}}_j^{\left( M \right)} = \left\{ {\begin{array}{*{20}{c}}
			{{\bf{0}},}&{j \ne u,}\\
			{{{\bf{H}}_{{\rm{eq}},u}}{\bf{\bar V}}_u^{\left( M \right)},}&{j = u.}
	\end{array}} \right.
	\tag{10}
\end{equation}

As \cite{Ni2016HybridBD} does, the SVD of ${{{\bf{H}}_{{\rm{eq}},u}}{\bf{\bar V}}_u^{\left( M \right)}}$ is ${{\bf{H}}_{{\rm{eq}},u}}{\bf{\bar V}}_u^{\left( M \right)} = {{\bf{U}}_u}{{\bf{\Sigma }}_u}{\bf{V}}_u^H$. The optimal digital precoder and combiner for the $u$-th user will be
\begin{equation}\label{equ:BD}
	{{\bf{F}}_{{\rm{BB}},u}} = {\bf{\bar V}}_u^{\left( M \right)}{\bf{V}}_u^{\left( {{N_{\rm{S}}}} \right)},{{\bf{W}}_{u,{\rm{opt}}}} = {\bf{U}}_u^{\left( {{N_{\rm{S}}}} \right)} ,
	\tag{11}
\end{equation}
where ${\bf{V}}_u^{\left( {{N_{\rm{S}}}} \right)}$ and ${\bf{U}}_u^{\left( {{N_{\rm{S}}}} \right)}$ are the first $N_{\rm{S}}$ column vectors of ${\bf{V}}_u$ and ${\bf{U}}_u$, respectively.

\subsection{Simplification of $R_u$}

According to \eqref{equ:null}, ${\bf {R}}_u$ in \eqref{equ:R_u} can be simplified as ${{\bf{R}}_u} = \sigma _u^2{\bf{W}}_u^H{{\bf{W}}_u}$ and $R_u$ in \eqref{equ:initial-Ru} can be rewritten as
\begin{equation}\label{equ:Ru1}
	\begin{array}{*{20}{l}}
		{\begin{array}{*{20}{c}}
				{{R_u} = {{\log }_2}\left| {{{\bf{I}}_{{N_{\rm{S}}}}} + \frac{\rho }{{\sigma _u^2U{N_{\rm{S}}}}}{{\left( {{\bf{W}}_u^H{{\bf{W}}_u}} \right)}^{ - 1}}{\bf{W}}_u^H{{\bf{H}}_u}} \right.}\\
				{\left. { \times {{\bf{F}}_{{\rm{RF}}}}{{\bf{F}}_{{\rm{BB}},u}}{\bf{F}}_{{\rm{BB}},u}^H{\bf{F}}_{{\rm{RF}}}^H{\bf{H}}_u^H{{\bf{W}}_u}} \right|} .
		\end{array}}
	\end{array}
	\tag{12}
\end{equation}

Substitute \eqref{equ:BD} into \eqref{equ:Ru1}, we have
\begin{equation}\label{equ:Ru2}
	{R_u} = {\log _2}\left| {{{\bf{I}}_{{N_{\rm{S}}}}} + \frac{{\rho {{\left( {{\bf{\Sigma }}_u^{\left( {{N_{\rm{S}}}} \right)}} \right)}^2}}}{{\sigma _u^2U{N_{\rm{S}}}}}} \right| .
	\tag{13}
\end{equation}
where ${{\bf{\Sigma }}_u^{\left( {{N_{\rm{S}}}} \right)}}$ is the first $ N_{\rm{S}} \times N_{\rm{S}}$ block of ${{\bf{\Sigma }}_u}$.

Besides, we can rewrite \eqref{equ:Ru1} as
\begin{equation}\label{equ:Ru3}
	\begin{array}{*{20}{c}}
		{{R_u} = {{\log }_2}\left| {{{\bf{I}}_{{N_{\rm{S}}}}} + \frac{\rho }{{\sigma _u^2U{N_{\rm{S}}}}}\sum\limits_{j = 1}^U {{{\left( {{\bf{W}}_u^H{{\bf{W}}_u}} \right)}^{ - 1}}{\bf{W}}_u^H{{\bf{H}}_u}} } \right.}\\
		{\left. { \times {{\bf{F}}_{{\rm{RF}},j}}{{\bf{F}}_{{\rm{BB}},u,j}}{\bf{F}}_{{\rm{BB}},u,j}^H{\bf{F}}_{{\rm{RF}},j}^H{\bf{H}}_u^H{{\bf{W}}_u}} \right|} ,
	\end{array}
	\tag{14}
\end{equation}
where ${{{\bf{F}}_{{\rm{BB}},u,j}}}$ means the $(j-1)M$ to $jM$ rows of ${{{\bf{F}}_{{\rm{BB}},u}}}$, $j = 1, \ldots ,U$.

Referring to the derivation in Section III.A, define ${{\bf{H}}_{{\rm{eq}},u,j}} = {{\bf{H}}_u}{{\bf{F}}_{{{\rm{RF}}},j}} \in {\mathbb{C}}^{N_{\rm{U}} \times M}$ and ${{\bf{\bar H}}_{u,j}} = {\left[ {{\bf{H}}_{{\rm{eq}},{u,1}}^T, \ldots ,{\bf{H}}_{{\rm{eq}},{u,j-1}}^T,{\bf{H}}_{{\rm{eq}},{u,j+1}}^T, \ldots ,{\bf{H}}_{{\rm{eq}},{u,U}}^T} \right]^T}$. The SVD of ${{\bf{\bar H}}_{u,j}}$ is ${{\bf{\bar H}}_{u,j}} = {{\bf{\bar U}}_{u,j}}{{\bf{\bar \Sigma }}_{u,j}}{\left[ {{\bf{\bar V}}_{u,j}^{\left( {\left( {U - 1} \right)M} \right)},{\bf{\bar V}}_{u,j}^{\left( M \right)}} \right]^H}$, where ${{\bf{\bar V}}_{u,j}^{\left( {\left( {U - 1} \right)M} \right)}}$ and ${{\bf{\bar V}}_{u,j}^{\left( M \right)}}$ are the first $(U-1)M$ and rest $M$ right singular vectors of ${{\bf{\bar H}}_{u,j}}$.

\begin{figure*}[ht] %hb代表放在文章底部，%ht为放在文章顶部
	%\hrulefill  %上面那条横线
	\begin{equation}\label{equ:convert-Ru}
		\begin{array}{*{20}{l}}
			{{{\tilde R}_u} = {{\log }_2}\left( {\left| {{{\bf{I}}_{{N_{\rm{S}}}}} + \frac{{\rho c}}{{\sigma _u^2U{N_{\rm{S}}}}}{\bf{W}}_u^\dag {{\bf{H}}_u}\left[ {{{\bf{F}}_{{\rm{RF}},u,M - 1}}\;{{\bf{f}}_{{\rm{RF}},u,M}}} \right]{{\left[ {{{\bf{F}}_{{\rm{RF}},u,M - 1}}\;{{\bf{f}}_{{\rm{RF}},u,M}}} \right]}^H}{\bf{H}}_u^H{{\bf{W}}_u}} \right|} \right)}\\
			{{\rm{ = }}{{\log }_2}\left( {\left| {{{\bf{I}}_{{N_{\rm{S}}}}} + \frac{{\rho c}}{{\sigma _u^2U{N_{\rm{S}}}}}{\bf{W}}_u^\dag {{\bf{H}}_u}{{\bf{F}}_{{\rm{RF}},u,M - 1}}{\bf{F}}_{{\rm{RF}},u,M - 1}^H{\bf{H}}_u^H{{\bf{W}}_u} + \frac{{\rho c}}{{\sigma _u^2U{N_{\rm{S}}}}}{\bf{W}}_u^\dag {{\bf{H}}_u}{{\bf{f}}_{{\rm{RF}},u,M}}{\bf{f}}_{{\rm{RF}},u,M}^H{\bf{H}}_u^H{{\bf{W}}_u}} \right|} \right)}\\
			{ = {{\log }_2}\left( {\left| {{{\bf{P}}_{u,M - 1}}} \right|} \right) + {{\log }_2}\left| {{{\bf{I}}_{{N_{\rm{S}}}}} + \frac{{\rho c}}{{\sigma _u^2U{N_{\rm{S}}}}}{\bf{P}}_{u,M - 1}^{ - 1}{\bf{W}}_u^\dag {{\bf{H}}_u}{{\bf{f}}_{{\rm{RF}},u,M}}{\bf{f}}_{{\rm{RF}},u,M}^H{\bf{H}}_u^H{{\bf{W}}_u}} \right|}\\
			{\mathop  = \limits^{\left( b \right)} {{\log }_2}\left( {\left| {{{\bf{P}}_{u,M - 1}}} \right|} \right) + {{\log }_2}\left| {1 + \frac{{\rho c}}{{\sigma _u^2U{N_{\rm{S}}}}}{\bf{f}}_{{\rm{RF}},u,M}^H{\bf{H}}_u^H{{\bf{W}}_u}{\bf{P}}_{u,M - 1}^{ - 1}{\bf{W}}_u^\dag {{\bf{H}}_u}{{\bf{f}}_{{\rm{RF}},u,M}}} \right|}
		\end{array}
		\tag {19}
	\end{equation}
	\hrulefill
\end{figure*}

Define the SVD of ${{{\bf{H}}_{{\rm{eq}},{u,j}}}{\bf{\bar V}}_{u,j}^{\left( M \right)}}$ is ${{\bf{H}}_{{\rm{eq}},{u,j}}}{\bf{\bar V}}_{u,j}^{\left( M \right)} = {{\bf{U}}_{u,j}}{{\bf{\Sigma }}_{u,j}}{\bf{V}}_{u,j}^H$. The optimal value of ${{{\bf{F}}_{{\rm{BB}},u,j}}}$ will be ${{\bf{F}}_{{\rm{BB}},{u,j}}} = {\bf{\bar V}}_{u,j}^{\left( M \right)}{\bf{V}}_{u,j}^{\left( {{N_{\rm{S}}}} \right)}$, where ${\bf{V}}_{u,j}^{\left( {{N_{\rm{S}}}} \right)}$ is the first $N_{\rm{S}}$ column vectors of ${\bf{V}}_{u,j}$. Substitute it into \eqref{equ:Ru1}, we have
\begin{equation}\label{equ:Ru4}
	\begin{array}{l}
		{R_u} = {\log _2}\left| {{{\bf{I}}_{{N_{\rm{S}}}}} + \frac{{\rho \sum\limits_{j = 1}^U {{{\left( {{\bf{\Sigma }}_{u,j}^{\left( {{N_{\rm{S}}}} \right)}} \right)}^2}} }}{{\sigma _u^2U{N_{\rm{S}}}}}} \right|\\
		\mathop  \ge \limits^{\left( a \right)} \sum\limits_{j = 1}^U {{{\log }_2}\left| {{{\bf{I}}_{{N_{\rm{S}}}}} + \frac{{\rho {{\left( {{\bf{\Sigma }}_{u,j}^{\left( {{N_{\rm{S}}}} \right)}} \right)}^2}}}{{\sigma _u^2U{N_{\rm{S}}}}}} \right|} ,
	\end{array}
	\tag{15}
\end{equation}
where ${{\bf{\Sigma }}_{u,j}^{\left( {{N_{\rm{S}}}} \right)}}$ is the first $ N_{\rm{S}} \times N_{\rm{S}}$ block of ${{\bf{\Sigma }}_{u,j}}$. $(a)$ is due to the Jensen's inequality and the equation holds when ${\bf{\Sigma }}_{u,1}^{\left( {{N_{\rm{S}}}} \right)} =  \ldots  = {\bf{\Sigma }}_{u,j}^{\left( {{N_{\rm{S}}}} \right)} =  \ldots  = {\bf{\Sigma }}_{u,U}^{\left( {{N_{\rm{S}}}} \right)}$. Then
\begin{equation}\label{}
	\begin{array}{l}
		\sum\limits_{j = 1}^U {{{\log }_2}\left| {{{\bf{I}}_{{N_{\rm{S}}}}} + \frac{{\rho {{\left( {{\bf{\Sigma }}_{u,j}^{\left( {{N_{\rm{S}}}} \right)}} \right)}^2}}}{{\sigma _u^2U{N_{\rm{S}}}}}} \right|}  = U{\log _2}\left| {{{\bf{I}}_{{N_{\rm{S}}}}} + \frac{{\rho {{\left( {{\bf{\Sigma }}_{u,u}^{\left( {{N_{\rm{S}}}} \right)}} \right)}^2}}}{{\sigma _u^2U{N_{\rm{S}}}}}} \right|\\
		= U\underbrace {\begin{array}{*{20}{c}}
				{{{\log }_2}\left| {{{\bf{I}}_{{N_{\rm{S}}}}} + \frac{\rho }{{\sigma _u^2U{N_{\rm{S}}}}}{{\left( {{\bf{W}}_u^H{{\bf{W}}_u}} \right)}^{ - 1}}{\bf{W}}_u^H{{\bf{H}}_u}} \right.}\\
				{\left. { \times {{\bf{F}}_{{\rm{RF}},u}}{{\bf{F}}_{{\rm{BB}},u,u}}{\bf{F}}_{{\rm{BB}},u,u}^H{\bf{F}}_{{\rm{RF}},u}^H{\bf{H}}_u^H{{\bf{W}}_u}} \right|}
		\end{array}}_{{{\tilde R}_u}}\\
		= U{{\tilde R}_u} ,
	\end{array}
	\tag{16}
\end{equation}
and $R_u \ge U{{\tilde R}_u}$. We can conclude that $U{{\tilde R}_u}$ implies a lower bound on $R_u$. The optimization problem of maximizing $R_u$ can be relaxed to the optimization problem of maximizing ${{\tilde R}_u}$.

Generally, when $M=N_{\rm{S}}$, ${\bf{F}}_{{\rm{BB}},{u,u}}$ is scaled unitary, i.e., ${{\bf{F}}_{{\rm{BB}},u,u}}{\bf{F}}_{{\rm{BB}},u,u}^H = {\bf{F}}_{{\rm{BB}},u,u}^H{{\bf{F}}_{{\rm{BB}},u,u}} = c{{\bf{I}}_M}$ \cite{Rusu2016LowCH}, where $c$ is a constant. Then
\begin{equation}\label{equ:Tidle_Ru1}
	{{\tilde R}_u} = {\log _2}\left| {{{\bf{I}}_{{N_{\rm{S}}}}} + \frac{{\rho c}}{{\sigma _u^2U{N_{\rm{S}}}}}{\bf{W}}_u^\dag {{\bf{H}}_u}{{\bf{F}}_{{\rm{RF}},u}}{\bf{F}}_{{\rm{RF}},u}^H{\bf{H}}_u^H{{\bf{W}}_u}} \right|.
	\tag{17}
\end{equation}
where $\mathbf{W}_{u}^{\dagger }={{\left( \mathbf{W}_{u}^{H}{{\mathbf{W}}_{u}} \right)}^{-1}}\mathbf{W}_{u}^{H}$ means the pseudo-inverse of ${{\mathbf{W}}_{u}}$. We can see that ${{\tilde R}_u}$ is simplified to a function mainly depending on ${{\bf{F}}_{{\rm{RF}},u}}$, which will help significantly simplify the design of analog precoder.

\subsection{Column-Wise Analog Precoding}

The solution of ${\bf{F}}_{{\rm{RF}},u}$ can be obtained by solving the following problem
\begin{equation}\label{equ:objective1}
  	\begin{array}{*{20}{c}}
		{{\bf{F}}_{{\rm{RF}},u}^{{\rm{opt}}} = \mathop {\arg \max }\limits_{{{\bf{F}}_{{\rm{RF}},u}}} {{\log }_2}\left| {{{\bf{I}}_{{N_{\rm{S}}}}} + \frac{{\rho c}}{{\sigma _u^2U{N_{\rm{S}}}}}{\bf{W}}_u^\dag {{\bf{H}}_u}} \right.}\\
		{\times \left. {{{\bf{F}}_{{\rm{RF}},u}}{\bf{F}}_{{\rm{RF}},u}^H{\bf{H}}_u^H{{\bf{W}}_u}} \right|}\\
		{s.t.\;\left| {{{\bf{F}}_{{\rm{RF}},u}}\left( {n,m} \right)} \right| = 1/\sqrt {{N_{{\rm{BS}}}}} .}
	\end{array}
\tag{18}
\end{equation}

Express ${\bf{F}}_{{\rm{RF}},u}$ as the horizontal concatenation of column vectors such that ${{\bf{F}}_{{\rm{RF}},u}} = \left[ {{{\bf{f}}_{{\rm{RF}},u,1}}, \ldots ,{{\bf{f}}_{{\rm{RF}},u,m}}, \ldots ,{{\bf{f}}_{{\rm{RF}},u,M}}} \right]$, where ${{\bf{f}}_{{\rm{RF}},u,m}}$ denotes the $m$-th column vector of ${{\bf{F}}_{{\rm{RF}},u}}$, we can convert \eqref{equ:Tidle_Ru1} to \eqref{equ:convert-Ru},
where ${{\mathbf{P}}_{u,M-1}}={{\mathbf{I}}_{{{N}_{\text{S}}}}}+\frac{\rho c}{\sigma _{u}^{2}U{{N}_{\text{S}}}}\mathbf{W}_{u}^{\dagger }{{\mathbf{H}}_{u}}{{\mathbf{F}}_{\text{RF},u,M-1}}\mathbf{F}_{\text{RF},u,M-1}^{H}\mathbf{H}_{u}^{H}{{\mathbf{W}}_{u}}$, ${{{\bf{F}}_{{\rm{RF}},u,M - 1}}}$ means the first $M-1$ column vectors of ${{{\bf{F}}_{{\rm{RF}},u}}}$, and the equation $\left( b \right)$ is due to $\left| {{\bf{I}} + {\bf{AB}}} \right| = \left| {{\bf{I}} + {\bf{BA}}} \right|$.

Let ${{\mathbf{P}}_{u,M-1}}=\mathbf{W}_{u}^{\dagger }{{\mathbf{Q}}_{u,M-1}}{{\mathbf{W}}_{u}}$, where ${{\mathbf{Q}}_{u,M-1}}=\left( {{\mathbf{I}}_{{{N}_{\text{U}}}}}+\frac{\rho c}{\sigma _{u}^{2}U{{N}_{\text{S}}}}{{\mathbf{H}}_{u}}{{\mathbf{F}}_{\text{RF},u,M-1}}\mathbf{F}_{\text{RF},u,M-1}^{H}\mathbf{H}_{u}^{H} \right)$, then
\begin{equation}\label{equ:Tidle_Ru2}
  \begin{array}{l}
{{\tilde R}_u} = {\log _2}\left( {\left| {{{\bf{P}}_{u,M - 1}}} \right|} \right) + \\
{\log _2}\left| {1 + \frac{{\rho c}}{{\sigma _u^2U{N_{\rm{S}}}}}{\bf{f}}_{{\rm{RF}},u,M}^H{\bf{H}}_u^H{\bf{Q}}_{u,M - 1}^{ - 1}{{\bf{H}}_u}{{\bf{f}}_{{\rm{RF}},u,M}}} \right|.
\end{array}
\tag {20}
\end{equation}

Similarly, by converting the first item of \eqref{equ:Tidle_Ru2}, ${{\tilde R}_u}$ can be further expressed as
\begin{equation}\label{}
  \begin{array}{l}
{{\tilde R}_u} = {\log _2}\left( {\left| {{{\bf{P}}_{u,M - 2}}} \right|} \right)\\
 + {\log _2}\left| {1 + \frac{{\rho c}}{{\sigma _u^2U{N_{\rm{S}}}}}{\bf{f}}_{{\rm{RF}},u,M - 1}^H{\bf{H}}_u^H{\bf{Q}}_{u,M - 2}^{ - 1}{{\bf{H}}_u}{{\bf{f}}_{{\rm{RF}},u,M - 1}}} \right|\\
 + {\log _2}\left| {1 + \frac{{\rho c}}{{\sigma _u^2U{N_{\rm{S}}}}}{\bf{f}}_{{\rm{RF}},u,M}^H{\bf{H}}_u^H{\bf{Q}}_{u,M - 1}^{ - 1}{{\bf{H}}_u}{{\bf{f}}_{{\rm{RF}},u,M}}} \right|,
\end{array}
\tag {21}
\end{equation}
where ${{\mathbf{P}}_{u,M-2}}=\mathbf{W}_{u}^{\dagger }{{\mathbf{Q}}_{u,M-2}}{{\mathbf{W}}_{u}}$, ${{\mathbf{Q}}_{u,M-2}}={{\mathbf{I}}_{{{N}_{\text{U}}}}}+\frac{\rho c}{\sigma _{u}^{2}U{{N}_{\text{S}}}}{{\mathbf{H}}_{u}}{{\mathbf{F}}_{\text{RF},u,M-2}}\mathbf{F}_{\text{RF},u,M-2}^{H}\mathbf{H}_{u}^{H}$, and ${{{\bf{F}}_{{\rm{RF}},u,M - 2}}}$ means the first $M-2$ column vectors of ${{{\bf{F}}_{{\rm{RF}},u}}}$.

By continuing converting the expression of ${{\tilde R}_u}$, eventually, ${{\tilde R}_u}$ can be computed by
\begin{equation}\label{}
  {{\tilde R}_u} = \sum\limits_{m = 1}^M {{{\log }_2}\left| {1 + \frac{{\rho c}}{{\sigma _u^2U{N_{\rm{S}}}}}{\bf{f}}_{{\rm{RF}},u,m}^H{\bf{H}}_u^H{\bf{Q}}_{u,m - 1}^{ - 1}{{\bf{H}}_u}{{\bf{f}}_{{\rm{RF}},u,m}}} \right|} ,
\tag {22}
\end{equation}
where ${{\mathbf{Q}}_{u,m-1}}={{\mathbf{I}}_{{{N}_{\text{U}}}}}+\frac{\rho c}{\sigma _{u}^{2}U{{N}_{\text{S}}}}{{\mathbf{H}}_{u}}{{\mathbf{F}}_{\text{RF},u,m-1}}\mathbf{F}_{\text{RF},u,m-1}^{H}\mathbf{H}_{u}^{H}$ ($m > 1$) and ${{{\bf{F}}_{{\rm{RF}},u,m - 1}}}$ means the first $m-1$ column vectors of ${{{\bf{F}}_{{\rm{RF}},u}}}$. The initial value ${{\mathbf{Q}}_{u,0}}$ is set to ${\bf{I}}_{N_{\rm{U}}}$.

Therefore, the optimization problem \eqref{equ:objective1} is transformed into a series of sub-rate optimization problems
\begin{equation}\label{equ:objective2}
\begin{array}{*{20}{c}}
{{\bf{f}}_{{\rm{RF}},u,m}^{{\rm{opt}}} = \mathop {{\rm{argmax}}}\limits_{{{\bf{f}}_{{\rm{RF}},u,m}}} {{\log }_2}\left| {1 + \frac{{\rho c}}{{{N_{\rm{S}}}\sigma _n^2}}{\bf{f}}_{{\rm{RF}},u,m}^H{\bf{H}}_u^H} \right.}\\
{ \times \left. {{\bf{Q}}_{u,m - 1}^{ - 1}{{\bf{H}}_u}{{\bf{f}}_{{\rm{RF}},u,m}}} \right|}\\
{s.t.\;\left| {{{\bf{f}}_{{\rm{RF}},u,m}}\left( n \right)} \right| = 1/\sqrt {{N_{{\rm{BS}}}}} .}
\end{array}
\tag {23}
\end{equation}

Through a series of derivations, the optimization problem \eqref{equ:objective1} is converted to the optimization problem \eqref{equ:objective2}. Optimizing ${{{\bf{F}}_{{\rm{RF}},u}}}$ in \eqref{equ:objective1} is transformed into optimizing each column vector of ${{{\bf{F}}_{{\rm{RF}},u}}}$ in \eqref{equ:objective2}, the solution of which is given by \cite{Gao2016EnergyHA}
\begin{equation}\label{}
  {\bf{f}}_{{\rm{RF}},u,m}^{{\rm{opt}}}\left( n \right) = \frac{1}{{\sqrt {{N_{{\rm{BS}}}}} }}{e^{j\angle {{\bf{v}}_{u,m}}\left( n \right)}} ,
\tag {24}
\end{equation}
where ${{\bf{v}}_{u,m}}$ is the right singular vector of $\mathbf{H}_{u}^{H}\mathbf{Q}_{u,m-1}^{-1}{{\mathbf{H}}_{u}}$ corresponding to the largest singular value and ${\angle {{\bf{v}}_{u,m}}\left( n \right)}$ means the phase angle of ${{\bf{v}}_{u,m}}\left( n \right)$.

\begin{figure}[bt!]
	\centering
	\includegraphics[width=3.5in]{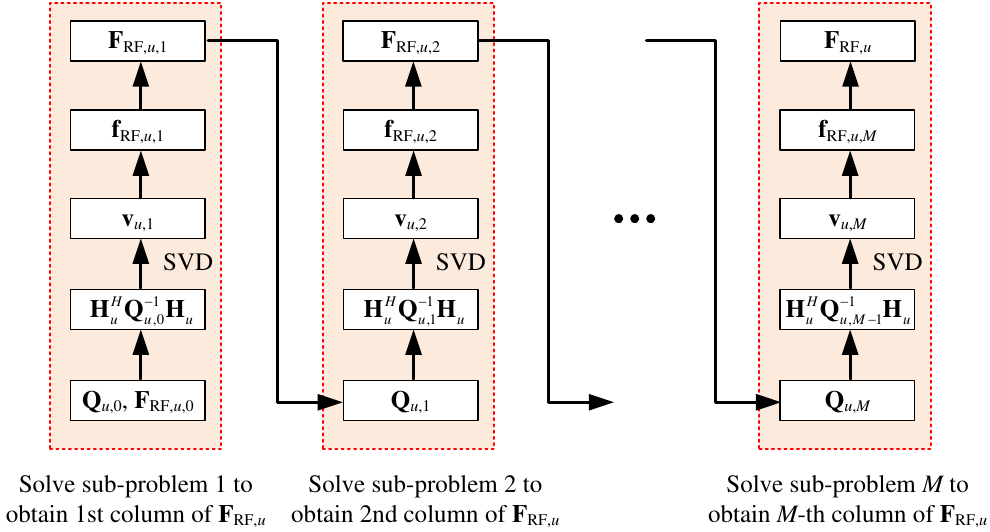}\\
	\caption{Illustration of the column-wise analog precoding for the $u$-th user.}
	\label{fig:fig-column-wise}
\end{figure}

Fig. \ref{fig:fig-column-wise} depicts the procedure of column-wise analog precoding for the $u$-th user. Each column of ${{{\bf{F}}_{{\rm{RF}},u}}}$ is computed in sequence, and eventually ${{{\bf{F}}_{{\rm{RF}},u}}}$ is obtained.

Considering the phase resolution of APSs, we denote the quantized bit of the phase as $B$. Each entry of the obtained analog precoder ${{\mathbf{F}}_{{\text{RF}},u}}$ will be quantized to the nearest value in the set ${{\mathcal{S}}_{\text{F}}}=\left\{ {{x}_{p}}\left| {{x}_{p}}=\frac{1}{\sqrt{{{N}_{\text{BS}}}}}{{e}^{j\frac{2\pi \left( p-1 \right)}{{{2}^{B}}}}},p=1,\ldots ,{{2}^{B}} \right. \right\}$ such that
\begin{equation}\label{equ:compute-FRF}
	{{{{\bf{\bar F}}}_{{{\rm{RF}}},u}}\left( {m,n} \right) = \mathop {\arg \min }\limits_{{x_p} \in {{\cal S}_{\rm{F}}}} \left| {{{\bf{F}}_{{{\rm{RF}}},u}}\left( {m,n} \right) - {x_p}} \right|} .
	\tag{25}
\end{equation}

After ${{\mathbf{\bar F}}_{{\text{RF}},u}}$ is obtained, the equivalent baseband channel is computed as ${{\bf{H}}_{{\rm{eq}},u}} = {{\bf{H}}_u}{{\bf{\bar F}}_{{\rm{RF}}}}$, where ${{\bf{\bar F}}_{{\rm{RF}}}} = \left[ {{{{\bf{\bar F}}}_{{\rm{RF,1}}}}, \ldots ,{{{\bf{\bar F}}}_{{\rm{RF,}}U}}} \right]$. According to \eqref{equ:Hu}, \eqref{equ:SVDHu} and \eqref{equ:BD}, the digital precoder ${{\mathbf{F}}_{\text{BB}}}$ and the fully-digital combiner for each user ${{\mathbf{W}}_{u,\text{opt}}}$ can be obtained. Then, ${{\mathbf{F}}_{\text{BB}}}$ is normalized to satisfy the power constraint
\begin{equation}\label{equ:compute-FBB}
  {{\bf{F}}_{{\rm{BB}}}} = {{\sqrt {U{N_{\rm{S}}}} {{\bf{F}}_{{\rm{BB}}}}} \mathord{\left/
 {\vphantom {{\sqrt {U{N_{\rm{S}}}} {{\bf{F}}_{{\rm{BB}}}}} {{{\left\| {{{\bf{F}}_{{\rm{RF}}}}{{\bf{F}}_{{\rm{BB}}}}} \right\|}_F}}}} \right.
 \kern-\nulldelimiterspace} {{{\left\| {{{\bf{F}}_{{\rm{RF}}}}{{\bf{F}}_{{\rm{BB}}}}} \right\|}_F}}} ,
\tag {26}
\end{equation}
and the hybrid combiner for each user is designed by using our previously-proposed AG algorithm \cite{Zhang2021FrequencySH}.

\subsection{AG Based Hybrid Combining}

\begin{algorithm}[bt!]
	\centering
	\caption{The Proposed CWAP-AGHC Scheme } \label{alg:CWAP}
	\begin{algorithmic} [1]
		\REQUIRE $c$, $T$, $\alpha$, $\varepsilon$, $\delta$;
		\STATE {\bf{I}. Column-Wise Analog Precoding}
		\FOR{$1 \le u \le U$}
		\STATE Initialize ${\bf{F}}_{{\rm{RF}},u} = \left[ \; \right]$, ${{\mathbf{Q}}_{u,0}}={{\mathbf{I}}_{{{N}_{\text{U}}}}}$;
		\FOR{$1 \le m \le M$}
		\IF{$m > 1$}
		\STATE ${{\mathbf{Q}}_{u,m-1}}={{\mathbf{I}}_{{{N}_{\text{U}}}}}+\frac{\rho c}{\sigma _{u}^{2}U{{N}_{\text{S}}}}{{\mathbf{H}}_{u}}{{\mathbf{F}}_{\text{RF},u,m-1}}\mathbf{F}_{\text{RF},u,m-1}^{H}\mathbf{H}_{u}^{H}$;
		\ENDIF
		\STATE Obtain ${\bf{v}}_{u,m}$ from the singular value decomposition of $\mathbf{H}_{u}^{H}\mathbf{Q}_{u,m-1}^{-1}{{\mathbf{H}}_{u}}$;
		\FOR{$1 \le n \le N_{\rm{U}}$}
		\STATE ${\bf{f}}_{{\rm{RF}},u,m}^{\rm{opt}}\left( n \right) = \frac{1}{{\sqrt {{N_{{\rm{BS}}}}} }}{e^{j\angle {{\bf{v}}_{u,m}}\left( n \right)}}$;
		\ENDFOR
		\ENDFOR
		\ENDFOR
		\STATE Quantize the analog precoder according to \eqref{equ:compute-FRF};
		\STATE Compute the equivalent baseband channel ${{\bf{H}}_{{\rm{eq}},u}} = {{\bf{H}}_u}{{\bf{\bar F}}_{{\rm{RF}}}}$;
		\STATE Apply BD method to obtain ${\bf{F}}_{\rm{BB}}^{\rm{opt}}$ and ${{\mathbf{W}}_{u,\text{opt}}}$ according to \eqref{equ:BD};
		\STATE Normalize ${{\mathbf{F}}_{\text{BB}}}$ according to \eqref{equ:compute-FBB};
		\STATE {\bf{II}. AG based hybrid combining}
		\FOR{$1 \le u \le U$}
		\STATE Initialize ${{\bf{W}}_{{\rm{RF}},u}}$ with random phases and the initialize ${{\bf{W}}_{{\rm{BB}},u}}$ according to \eqref{equ:compute-WBB};
		\STATE Obtain the vectors ${{\mathbf{x}}_{u}}=vec({{\mathbf{W}}_{\text{RF},u}})$ and ${{\mathbf{y}}_{u}}=vec({{\mathbf{W}}_{u,\text{opt}}})$;
		\FOR{$1 \le t \le T$}
		\IF{$\left\| {{\mathbf{W}}_{u,\text{opt}}}-{{\mathbf{W}}_{\text{RF},u}}{{\mathbf{W}}_{\text{BB},u}} \right\|_{F}^{2}>\delta $}
		\STATE ${\bf{A}}_u = {{{\bf{W}}_{{\rm{BB}},u}}^T \otimes {{\bf{I}}_{{N_{\rm{U}}}}}}$;
		\STATE ${{\mathbf{w}}_{u}}={{\mathbf{y}}_{u}}-{{\mathbf{A}}_{u}}{{\mathbf{x}}_{u}}$;
		\FOR{$1 \le k \le {N_{\rm{U}}}M$}
		\STATE Compute the first-order derivative ${d_{u,k}}(t) = - 2{\bf{A}}_u{\left( {:,k} \right)^H} {\bf{w}}_u$;
		\STATE Update ${{x}_{u,k}}$ according to \eqref{equ:compute-xuk} and reconstruct ${{x}_{u,k}}$ to satisfy the constant-modulus constraint ${x_{u,k}} = \frac{1}{{\sqrt {{N_{\rm{U}}}} }}{e^{j\angle {x_{u,k}}}}$;
		\ENDFOR
		\STATE Reshape ${{\mathbf{x}}_{u}}$ to form ${{\bf{W}}_{{\rm{RF}},u}} = \mathcal{R}\left( {\bf{x}}_u \right)$;
		\STATE Compute ${{\bf{W}}_{{\rm{BB}},u}}$ according to \eqref{equ:compute-WBB}.
		\ENDIF
		\ENDFOR
		\ENDFOR
		\STATE Quantize the analog combiner of each user according to \eqref{equ:compute-WRF};
		\RETURN ${{\bf{\bar F}}_{{\rm{RF}}}^{\rm{opt}}}$, ${{\bf{F}}_{{\rm{BB}}}^{\rm{opt}}}$, ${{\bf{\bar W}}_{{\rm{RF}},u}^{\rm{opt}}}$, ${{\bf{W}}_{{\rm{BB}},u}^{\rm{opt}}}$.
	\end{algorithmic}
\end{algorithm}

For the $u$-th user, the design of ${\bf{W}}_{{\rm{RF}},u}$ and ${\bf{W}}_{{\rm{BB}},u}$ is to minimize the Euclidean distance between ${{\mathbf{W}}_{u,\text{opt}}}$ and ${\bf{W}}_{{\rm{RF}},u}{\bf{W}}_{{\rm{BB}},u}$ with constant-modulus constraint of ${\bf{W}}_{{\rm{RF}},u}$
\begin{equation}\label{equ:objective3}
  	\begin{array}{*{20}{c}}
		{\left( {{\bf{W}}_{{\rm{RF}},u}^{{\rm{opt}}},{\bf{W}}_{{\rm{BB}},u}^{{\rm{opt}}}} \right) = } \\
		{\mathop {{\rm{argmin}}}\limits_{{{\bf{W}}_{{\rm{RF}},u}},{{\bf{W}}_{{\rm{BB}},u}}} \left\| {{{\bf{W}}_{u,{\rm{opt}}}} - {{\bf{W}}_{{\rm{RF}},u}}{{\bf{W}}_{{\rm{BB}},u}}} \right\|_F^2}\\
		{s.t.\;\;\left| {{{\bf{W}}_{{\rm{RF}},u}}\left( {n',m'} \right)} \right| = 1/\sqrt {{N_{\rm{U}}}} .}
	\end{array}
\tag {27}
\end{equation}

By vectorizing ${\bf{W}}_{{\rm{RF}},u}$ and ${{\mathbf{W}}_{u,\text{opt}}}$, we obtain two vectors, ${\bf{x}}_u$ and ${\bf{y}}_u$ such that ${\bf{x}}_u = vec({\bf{W}}_{{\rm{RF}},u}) \in \mathbb{C}^{N_{\rm{U}}M \times 1}$ and ${{\mathbf{y}}_{u}}=vec({{\mathbf{W}}_{u,\text{opt}}})\in {{\mathbb{C}}^{{{N}_{\text{U}}}{{N}_{\text{S}}}\times 1}}$. The objective function in \eqref{equ:objective3} can be regarded as a function of ${\bf{x}}_u$, i.e., $f({\bf{x}}_u)$, and expressed as the following form
\begin{equation}\label{}
  \begin{array}{*{20}{l}}
\begin{array}{l}
f\left( {{{\bf{x}}_u}} \right) = \left\| {{{\bf{W}}_{u,{\rm{opt}}}} - {{\bf{W}}_{{\rm{RF}},u}}{{\bf{W}}_{{\rm{BB}},u}}} \right\|_F^2\\
 = \left\| {vec\left( {{{\bf{W}}_{u,{\rm{opt}}}}} \right) - \left( {{{\bf{W}}_{{\rm{BB}},u}}^T \otimes {{\bf{I}}_{{N_{\rm{U}}}}}} \right)vec\left( {{{\bf{W}}_{{\rm{RF}},u}}} \right)} \right\|_2^2
\end{array}\\
{ = \left\| {{{\bf{y}}_u} - {{\bf{A}}_u}{{\bf{x}}_u}} \right\|_2^2 = \sum\limits_{i = 1}^{{N_{\rm{U}}}{N_{\rm{S}}}} {{{\left| {{y_{u,i}} - {{\bf{A}}_u}\left( {i,:} \right){{\bf{x}}_u}} \right|}^2}} ,}
\end{array}
\tag {28}
\end{equation}
where ${\bf{A}}_u = {{\bf{W}}_{{\rm{BB}},u}}^T \otimes {{\bf{I}}_{{N_{\rm{U}}}}}$ is introduced for simplicity.

The first-order derivative of $f({\bf{x}}_u)$ respect to the $k$-th element of ${\bf{x}}_u$ (denoted as $x_{u,k}$) is computed by
\begin{equation}\label{}
  \begin{array}{l}
{d_{u,k}} = \frac{{\partial f\left( {{{\bf{x}}_u}} \right)}}{{\partial {x_{u,k}}}}\\
 =  - 2\sum\limits_{i = 1}^{{N_{\rm{U}}}{N_{\rm{S}}}} {{{\bf{A}}_u}{{\left( {i,k} \right)}^*}\left( {{y_{u,i}} - {{\bf{A}}_u}\left( {i,:} \right){{\bf{x}}_u}} \right)} \\
 =  - 2{{\bf{A}}_u}{\left( {:,k} \right)^H}\left( {{{\bf{y}}_u} - {{\bf{A}}_u}{{\bf{x}}_u}} \right) .
\end{array}
\tag {29}
\end{equation}

\begin{table*}[bt!]
	\centering
	\renewcommand{\arraystretch}{1.2}
	\caption{Numbers of complex multiplications and divisions of the proposed algorithms and the competing algorithm}
	\label{table:complexity}
	\begin{tabular}{|c|c|}
		\hline
		{\bf Algorithm} & {\bf Number of complex multiplications and divisions} \\
		\hline
		PE-AltMin & $N_{iter,\text{F}}^{\text{PE}}\left( 2{{N}_{\text{BS}}}{{U}^{2}}M{{N}_{\text{S}}}+{{N}_{\text{BS}}}UM+{{U}^{3}}M{{N}_{\text{S}}}^{2} \right)+N_{iter,\text{W}}^{\text{PE}}\left( 2{{N}_{\text{U}}}M{{N}_{\text{S}}}+{{N}_{\text{U}}}M+M{{N}_{\text{S}}}^{2} \right)$ \\
		\hline
		Proposed CWAP-AGHC & $\begin{matrix}
			U\left( \frac{M\left( M-1 \right)}{2}{{N}_{\text{BS}}}{{N}_{\text{U}}}+\frac{M\left( M-1 \right)\left( 2M-1 \right)}{6}{{N}_{\text{U}}}^{2}+{{N}_{\text{BS}}}{{N}_{\text{U}}}^{2}+{{N}_{\text{BS}}}^{2}{{N}_{\text{U}}}+{{f}_{\text{inv}}}\left( {{N}_{\text{U}}} \right)+2{{N}_{\text{U}}} \right)  \\
			+N_{iter,\text{W}}^{\text{AG}}\left( 2{{N}_{\text{U}}}^{2}M{{N}_{\text{S}}}+5{{N}_{\text{U}}}M+2{{N}_{\text{U}}}{{M}^{2}}+{{N}_{\text{U}}}M{{N}_{\text{S}}}+{{f}_{\text{inv}}}\left( M \right) \right)  \\
		\end{matrix}$ \\
		\hline
	\end{tabular}
\end{table*}

\begin{table*}[bt!]
	\caption{Parameters setting}
	\label{table:parameters}
	\centering
	\renewcommand{\arraystretch}{1.2}
	\begin{tabular}{|c|c|c|c|c|}
		\hline
		\bf Meaning & \bf Notation & \multicolumn{3}{|c|}{\bf Value} \\
		\hline
		Signal to noise ratio (SNR) & - & \multicolumn{3}{|c|}{$-20\sim20$ dB} \\
		\hline
		Number of antennas at the BS & ${{N}_{\text{BS}}}$ & \multicolumn{3}{|c|}{64, 128, 256, 512, 1024} \\
		\hline
		Number of users	& $U$ & \multicolumn{3}{|c|}{$2\sim8$} \\
		\hline
		Number of antennas at each user	& ${{N}_{\text{U}}}$ & 1, 2 & 4 & 8, 16 \\
		\hline
		Number of RF chains at each user & $M$ & 1 & 2 & 4 \\
		\hline
		Number of data streams at each user & ${{N}_{\text{S}}}$ & 1 & 1 or 2 & 2 or 4 \\
		\hline
		Resolution of APSs & $B$ & \multicolumn{3}{|c|}{1, 2, 3} \\
		\hline
		Number of propagation paths & ${{N}_{\text{p}}}$ & \multicolumn{3}{|c|}{5} \\
		\hline
		Complex gain of each propagation path & $\alpha_l$ & \multicolumn{3}{|c|}{Complex Gaussian distribution with mean $0$ and standard deviation $1$} \\
		\hline
		Azimuth angle of each propagation path & ${\phi}_{u,{\rm{T}}}^l$, ${\phi}_{u,{\rm{R}}}^l$ & \multicolumn{3}{|c|}{Uniformly distributed in $[0,2\pi)$} \\
		\hline
		Elevation angle of each propagation path & ${\theta}_{u,{\rm{T}}}^l$, ${\theta}_{u,{\rm{R}}}^l$ & \multicolumn{3}{|c|}{Uniformly distributed in $[0,\pi )$} \\
		\hline
		Maximum iterations & $T$ & \multicolumn{3}{|c|}{200} \\
		\hline
		Stop threshold of iterations & $\delta$ & \multicolumn{3}{|c|}{${{10}^{-2}}$} \\
		\hline
		Global step size of the proposed algorithm & $\alpha$ & \multicolumn{3}{|c|}{0.9} \\
		\hline
		\multirow{2}{*}{Other parameters of the proposed algorithm} & $\varepsilon$ & \multicolumn{3}{|c|}{$10^{-6}$} \\
		\cline{2-5}
		& $c$ & \multicolumn{3}{|c|}{1} \\
		\hline
	\end{tabular}
\end{table*}

To minimize $f \left( {\bf{x}}_u \right)$, the AG algorithm \cite{Duchi2011AdaptiveSM} is adopted to search the next point of $x_{u,k}$ along the negative gradient direction and the step size is related to the gradients of all the previous iterations
\begin{equation}\label{equ:compute-xuk}
  {x_{u,k}}\left( t \right) = {x_{u,k}}\left( {t - 1} \right) - \frac{\alpha }{{\sqrt {\sum\limits_{\tau  = 1}^t {{{\left| {{d_{u,k}}\left( \tau  \right)} \right|}^2}}  + \varepsilon } }}{d_{u,k}}\left( t \right) ,
\tag {30}
\end{equation}
where $t$ is the number of iterations, $\alpha$ is the global step size and $\varepsilon > 0$ is a sufficiently small value.
Considering the constant-modulus constraint imposed on $x_{u,k}$, we reconstruct $x_{u,k}$ by ${x_{u,k}} = \frac{1}{{\sqrt {{N_{\rm{U}}}} }}{e^{j\angle {x_{u,k}}}}$, and then reshape ${\bf{x}}_u$ to an $N_{\rm{U}} \times M$ matrix such that ${{\bf{W}}_{{\rm{RF}},u}} = \mathcal{R}\left( {\bf{x}}_u \right)$, where $\mathcal{R}\left( \cdot \right):{\mathbb{C}^{N_{\rm{U}}M \times 1}} \to {\mathbb{C}^{{N_{\rm{U}}} \times M}}$ is the reshaping function.

After ${{\bf{W}}_{{\rm{RF}},u}}$ is updated, ${{\bf{W}}_{{\rm{BB}},u}}$ is computed according to the least square solution
\begin{equation}\label{equ:compute-WBB}
  {{\bf{W}}_{{\rm{BB}},u}} = {\left( {{\bf{W}}_{{\rm{RF}},u}^H{{\bf{W}}_{{\rm{RF}},u}}} \right)^{ - 1}}{\bf{W}}_{{\rm{RF}},u}^H{\bf{W}}_{u,{{\rm{opt}}}} .
\tag {31}
\end{equation}

${{\bf{W}}_{{\rm{RF}},u}}$ and ${{\bf{W}}_{{\rm{BB}},u}}$ are updated through iterations until $\left\| {{\mathbf{W}}_{u,\text{opt}}}-{{\mathbf{W}}_{\text{RF},u}}{{\mathbf{W}}_{\text{BB},u}} \right\|_{F}^{2}$ is smaller than a sufficiently small positive value, i.e., $\delta$, or the number of iterations exceeds a maximum, i.e., $T$.

Each entry of the obtained analog combiners ${\bf{W}}_{{\rm{RF}},u}$ will be quantized to the nearest value in the set ${{\mathcal{S}}_{\text{W}}}=\left\{ {{x}_{q}}\left| {{x}_{q}}=\frac{1}{\sqrt{{{N}_{\text{U}}}}}{{e}^{j\frac{2\pi \left( q-1 \right)}{{{2}^{B}}}}},q=1,\ldots ,{{2}^{B}} \right. \right\}$ such that
\begin{equation}\label{equ:compute-WRF}
	{{{{\bf{\bar W}}}_{{\rm{RF}},u}}\left( {m',n'} \right) = \mathop {\arg \min }\limits_{{x_q} \in {{\cal S}_{\rm{W}}}} \left| {{{\bf{W}}_{{\rm{RF}},u}}\left( {m',n'} \right) - {x_q}} \right|} .
	\tag{32}
\end{equation}

The proposed column-wise analog precoding and AG based hybrid combing (CWAP-AGHC) scheme is summarized in {\bf {Algorithm \ref{alg:CWAP}}}.

\subsection{Complexity Analysis and Comparison}

The number of complex multiplications and divisions is used to evaluate the computational complexity of the proposed algorithms and the competing algorithms. The computational complexity of {\bf {Algorithm \ref{alg:CWAP}}} mainly comes from Steps 6, 8, 10, 24, 26, 27 and 30. The numbers of complex multiplications/divisions at these steps are ${{N}_{\text{BS}}}{{N}_{\text{U}}}\left( m-1 \right)+{{N}_{\text{U}}}^{2}{{\left( m-1 \right)}^{2}}$, ${{N}_{\text{BS}}}{{N}_{\text{U}}}^{2}+{{N}_{\text{BS}}}^{2}{{N}_{\text{U}}}+{{f}_{\text{inv}}}\left( {{N}_{\text{U}}} \right)$, 2, ${{N}_{\text{BS}}}^{2}{{U}^{2}}M{{N}_{\text{S}}}$, ${{N}_{\text{BS}}}UM$, 5 and $2{{N}_{\text{BS}}}{{U}^{2}}{{M}^{2}}+{{N}_{\text{BS}}}{{U}^{2}}M{{N}_{\text{S}}}+{{f}_{\text{inv}}}\left( UM \right)$, where ${f_{{\rm{inv}}}}\left( N \right) = {{\left( {{N^3} - {N^2} + N} \right)} \mathord{\left/
		{\vphantom {{\left( {{N^3} - {N^2} + N} \right)} 3}} \right.
		\kern-\nulldelimiterspace} 3}$
means the number of multiplications/divisions of inverting an $N\times N$ matrix through elementary row transformation. Considering the numbers of iterations and cycles, the total computational complexity of {\bf {Algorithm \ref{alg:CWAP}}} is listed in Table \ref{table:complexity}, where $N_{iter,\text{W}}^{\text{AG}}$ means the total number of iterations of all the users to obtain the hybrid combiners by using the AG algorithm.

The numbers of complex multiplications and divisions of the proposed algorithms and the competing algorithms are listed in Table \ref{table:complexity}, where $N_{iter,\text{W}}^{\text{PE}}$ and $N_{iter,\text{W}}^{\text{PE}}$ denote the numbers of iterations to design the hybrid precoder and hybrid combiners of all the users by using the PE-AltMin algorithm. For the kind of AltMin algorithms, the fully-digital precoder and fully-digital combiners of all the users are computed by using the BD method at first. Then the corresponding AltMin algorithm is executed to design the hybrid precoder and hybrid combiners of all the users. Because all the AltMin algorithms can achieve approximately same sum-rate, among which the PE-AltMin algorithm has the lowest complexity \cite{Zhang2021MachineLB}. So only the numbers of complex multiplications and divisions of the PE-AltMin algorithm is compared with that of the proposed CWAP-AGHC scheme. For the same reason, only the achievable sum-rate and complexity of the PE-AltMin algorithm will be compared with those of the proposed CWAP-AGHC scheme in Section \ref{sec-sim}.

\section{Simulation Results} \label{sec-sim}

The performance of the proposed algorithm is evaluated in this section. The parameters setting is listed in Table \ref{table:parameters}. One thousand channel examples are randomly generated according to \eqref{equ:channel} and \eqref{equ:array-response-vector}. For each channel example, the hybrid precoders and combiners are computed by using the PE-AltMin algorithm \cite{Yu2016AlternatingMA} and the proposed CWAP-AGHC scheme. The average results over one thousand channel examples are given subsequently.

\begin{figure} [bt!]
	\centering
	\subfloat[]{
		\includegraphics[width=3in]{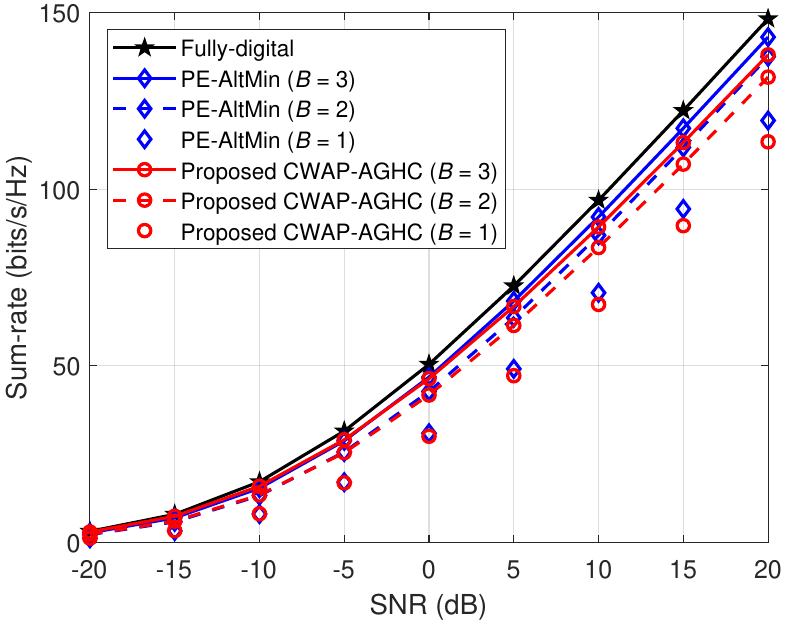}} \\
	\subfloat[]{
		\includegraphics[width=3in]{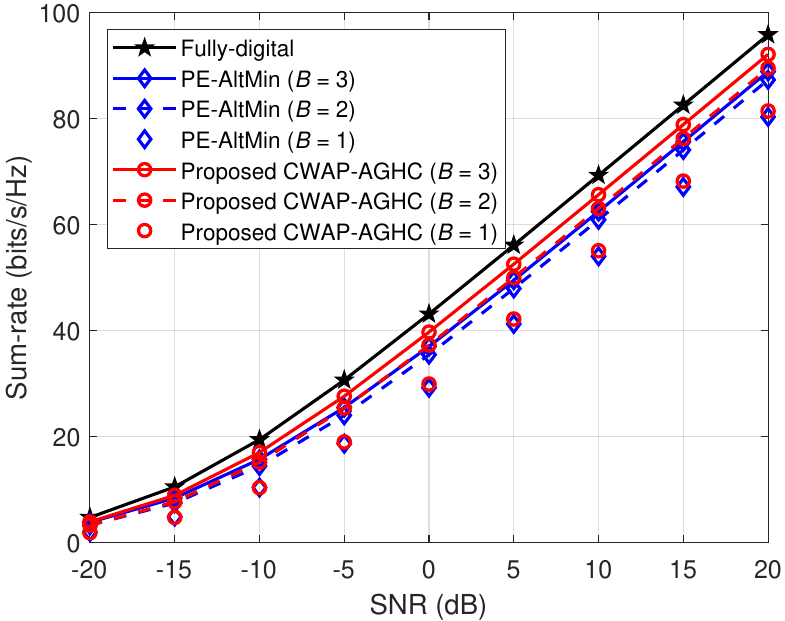}}
	\caption{Sum-rate against SNR when ${{N}_{\text{BS}}}=256$, ${{N}_{\text{U}}}=16$ and $U=4$. (a) The case of $M=N_{\rm{S}}$, (b) The case of $M>N_{\rm{S}}$.}
	\label{fig:sim-SE-SNR}
\end{figure}

\begin{figure} [bt!]
	\centering
	\subfloat[]{
		\includegraphics[width=3in]{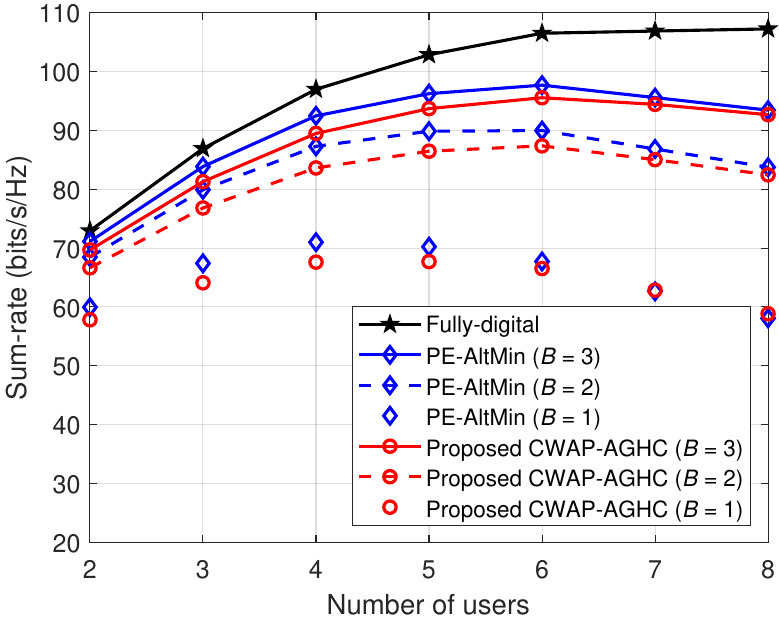}} \\
	\subfloat[]{
		\includegraphics[width=3in]{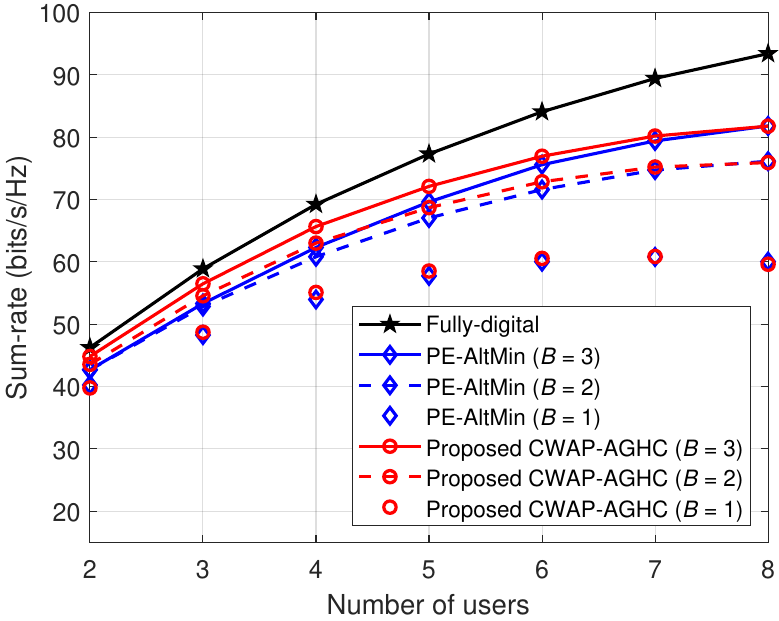}}
	\caption{Sum-rate against $U$ when ${{N}_{\text{BS}}}=256$, ${{N}_{\text{U}}}=16$ and SNR = 10 dB. (a) The case of $M=N_{\rm{S}}$, (b) The case of $M>N_{\rm{S}}$.}
	\label{fig:sim-SE-U}
\end{figure}

\begin{table*}[bt!]
	\centering
	\renewcommand{\arraystretch}{1.2}
	\caption{Numerical complex multiplications and divisions of the proposed and the competing algorithm}
	\label{table:complex-number-SNR}
	\begin{tabular}{|c|c|c|}
		\hline
		{\bf Algorithm} & {\bf Average number of iterations} & {\bf Number of complex multiplications and divisions} \\
		\hline
		PE-AltMin ($M=N_{\rm{S}}$) & $\bar{N}_{iter,\text{F}}^{\text{PE}}=172.10,\bar{N}_{iter,\text{W}}^{\text{PE}}=66.43$ & $2.40 \times {10^7}$ \\
		\hline
		Proposed CWAP-AGHC ($M=N_{\rm{S}}$) & $\bar{N}_{iter,\text{W}}^{\text{AG}}=51.60$ & $6.23 \times {10^6}$ \\
		\hline
		PE-AltMin ($M>N_{\rm{S}}$) & $\bar{N}_{iter,\text{F}}^{\text{PE}}=116.77,\bar{N}_{iter,\text{W}}^{\text{PE}}=44.80$ & $8.27 \times {10^6}$ \\
		\hline
		Proposed CWAP-AGHC ($M>N_{\rm{S}}$) & $\bar{N}_{iter,\text{W}}^{\text{AG}}=29.42$ & $4.73 \times {10^6}$ \\
		\hline
	\end{tabular}
\end{table*}

\begin{figure} [bt!]
	\centering
	\subfloat[]{
		\includegraphics[width=3in]{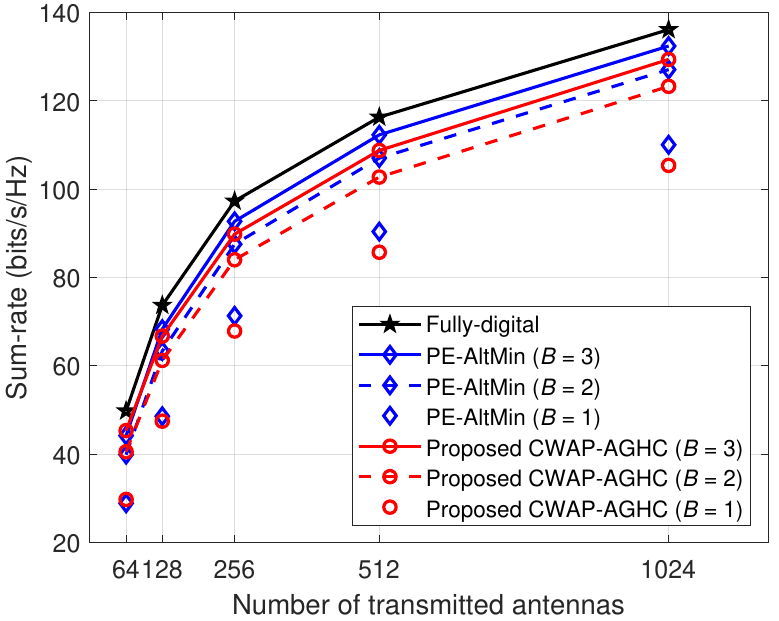}} \\
	\subfloat[]{
		\includegraphics[width=3in]{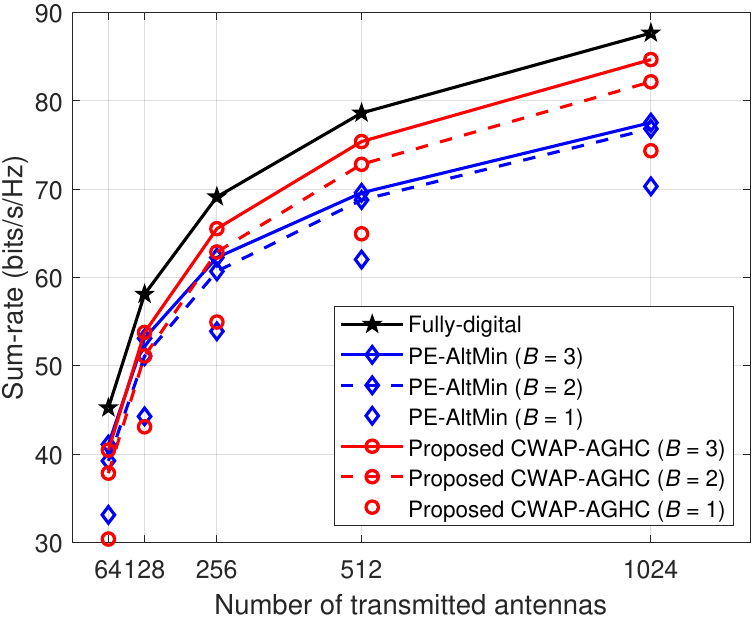}}
	\caption{Sum-rate against ${{N}_{\text{BS}}}$ when ${{N}_{\text{U}}}=16$, $U=4$ and SNR = 10 dB. (a) The case of $M=N_{\rm{S}}$, (b) The case of $M>N_{\rm{S}}$.}
	\label{fig:sim-SE-Nt}
\end{figure}

\begin{figure} [bt!]
	\centering
	\subfloat[]{
		\includegraphics[width=3in]{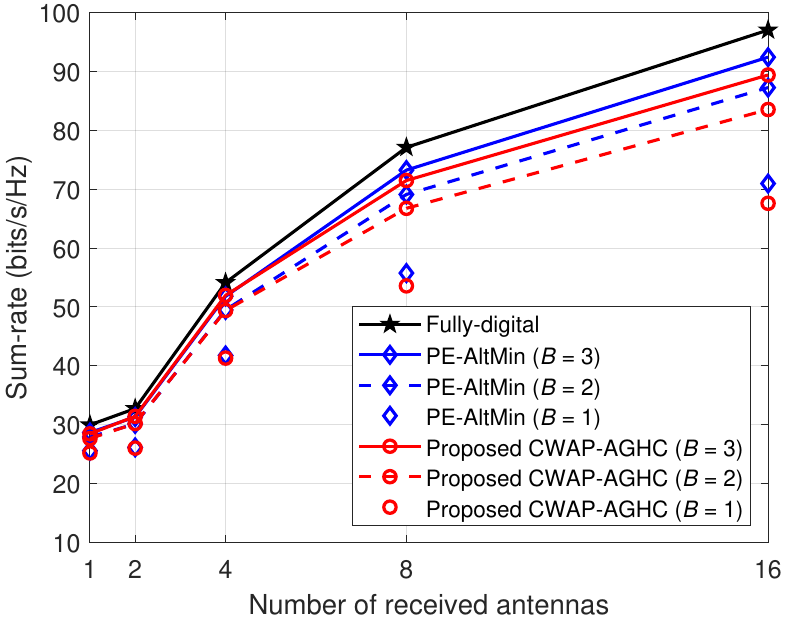}} \\
	\subfloat[]{
		\includegraphics[width=3in]{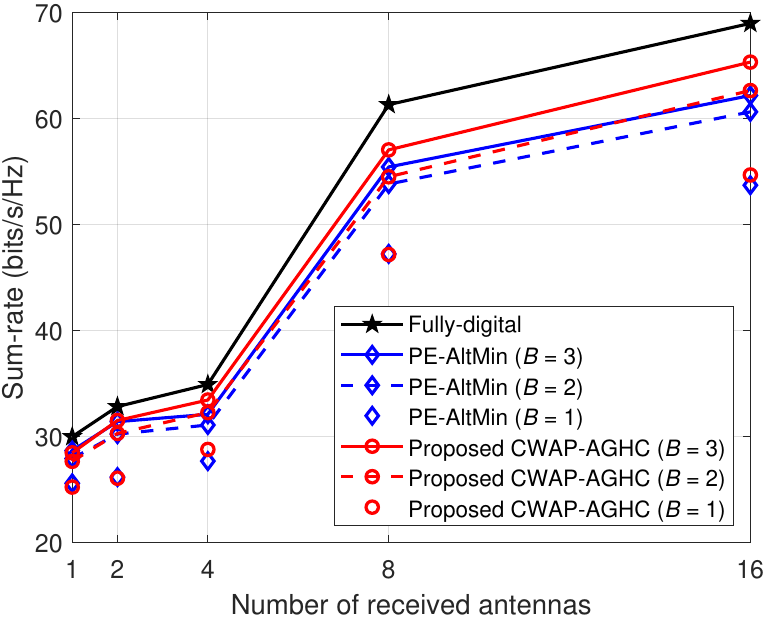}}
	\caption{Sum-rate against ${{N}_{\text{U}}}$ when ${{N}_{\text{BS}}}=256$, $U=4$ and SNR = 10 dB. (a) The case of $M=N_{\rm{S}}$, (b) The case of $M>N_{\rm{S}}$.}
	\label{fig:sim-SE-Nr}
\end{figure}

Generally, the number of RF chains $M$ is no less than the number of dara streams $N_{\rm{S}}$ for each user. We will consider two cases, i.e., $M=N_{\rm{S}}$ and $M>N_{\rm{S}}$, and show the simulation results in these two cases, respectively.

Fig. \ref{fig:sim-SE-SNR} shows the achievable sum-rate against SNR given ${{N}_{\text{BS}}}=256$, ${{N}_{\text{U}}}=16$ and $U=4$ when (a) $M=N_{\rm{S}}$ and (b) $M>N_{\rm{S}}$. As SNR increases, the sum-rate of all the algorithms increase. When $M=N_{\rm{S}}=4$, the proposed CWAP-AGHC scheme can achieve approximately same sum-rate as the PE-AltMin algorithm. When $M=4$ and $N_{\rm{S}}=2$, the CWAP-AGHC scheme outperforms the PE-AltMin algorithm in terms of the sum-rate. We can also conclude that, even with relatively-low-resolution APSs, i.e., $B=3$, the achievable sum-rate of the proposed CWAP-AGHC scheme can approach that of the fully-digital precoding and combining.

The numerical values of required iterations of all the algorithms are obtained through simulation, as listed in Table \ref{table:complex-number-SNR}. Substituting these values into corresponding formulas in Table \ref{table:complexity}, the number of complex multiplications and divisions can be obtained. The complexity of the proposed CWAP-AGHC scheme is lower than that of the PE-AltMin algorithm because the analog precoder is easily obtained by the closed-form solution instead of iterations. Since the PE-AltMin algorithm has lower complexity than MO-AltMin, BB-AltMin and GP-AltMin algorithms, we can conclude that the proposed CWAP-AGHC scheme has lower complexity than the AltMin algorithms.

Fig. \ref{fig:sim-SE-U} depicts the achievable sum-rate versus the number of users $U$ given ${{N}_{\text{BS}}}=256$, ${{N}_{\text{U}}}=16$ and SNR = 10 dB. When $M=N_{\rm{S}}=4$, the sum-rate of the proposed CWAP-AGHC scheme can approach that of the PE-AltMin algorithm. When $M=4$ and $N_{\rm{S}}=2$, the sum-rate of the proposed CWAP-AGHC scheme is slightly higher than that of the PE-AltMin algorithm. The number of complex multiplications and divisions with $U$ is shown in Table \ref{table:complex-number-SNR-U}. When $U\ge 4$, complexity of the proposed CWAP-AGHC scheme is much lower than that of the PE-AltMin algorithm. Almost one order of magnitude reduction of the number of complex multiplications and divisions is shown when $U=8$.

\begin{table*}[bt!]
	\centering
	\renewcommand{\arraystretch}{1.3}
	\caption{Numerical complex multiplications and divisions of the proposed and the competing algorithm against $U$}
	\label{table:complex-number-SNR-U}
	\begin{tabular}{|c|c|c|c|c|c|c|c|c|}
		\hline
		\multicolumn{2}{|c|}{$U$} & 2 &3 & 4 & 5 & 6 & 7 & 8 \\
		\hline
		\multirow{3}{*}{PE-AltMin ($M=N_{\rm{S}}$)} & $\bar{N}_{iter,\text{F}}^{\text{PE}}$ & 41.43 & 94.87 & 174.15 & 188.37 & 179.00 & 185.38 & 195.83 \\
		\cline{2-9}
		& $\bar{N}_{iter,\text{W}}^{\text{PE}}$ & 31.31 & 48.41 & 65.28 & 80.17 & 96.65 & 113.44 & 128.73 \\
		\cline{2-9}
		& $N_{\rm{Mul}}$ & $1.48 \times 10^6$ & $7.48 \times 10^6$ & $2.43 \times 10^7$ & $4.11 \times 10^7$ & $5.64 \times 10^7$ & $7.99 \times 10^7$ & $1.11 \times 10^8$ \\
		\hline
		\multirow{2}{*}{CWAP-AGHC ($M=N_{\rm{S}}$)} & $\bar{N}_{iter,\text{W}}^{\text{AG}}$ & 26.51 & 39.80 & 52.38 & 66.54 & 79.48 & 92.32 & 106.00 \\
		\cline{2-9}
		& $N_{\rm{Mul}}$ & $2.54 \times 10^6$ & $3.81 \times 10^6$ & $5.07 \times 10^6$ & $6.35 \times 10^6$ & $7.62 \times 10^6$ & $8.88 \times 10^6$ & $1.02 \times 10^7$ \\
		\hline
		\multirow{3}{*}{PE-AltMin ($M>N_{\rm{S}}$)} & $\bar{N}_{iter,\text{F}}^{\text{PE}}$ & 63.33 & 94.73 & 122.32 & 129.58 & 146.70 & 158.74 & 176.29 \\
		\cline{2-9}
		& $\bar{N}_{iter,\text{W}}^{\text{PE}}$ & 23.28 & 33.83 & 45.09 & 56.60 & 68.01 & 78.21 & 87.91 \\
		\cline{2-9}
		& $N_{\rm{Mul}}$ & $1.18 \times 10^6$ & $3.84 \times 10^6$ & $8.66 \times 10^6$ & $1.42 \times 10^7$ & $2.31 \times 10^7$ & $3.39 \times 10^7$ & $4.91 \times 10^7$ \\
		\hline
		\multirow{2}{*}{CWAP-AGHC ($M>N_{\rm{S}}$)} & $\bar{N}_{iter,\text{W}}^{\text{AG}}$ & 14.71 & 21.85 & 29.27 & 36.42 & 43.63 & 51.11 & 58.23 \\
		\cline{2-9}
		& $N_{\rm{Mul}}$ & $2.37 \times 10^6$ & $3.55 \times 10^6$ & $4.73 \times 10^6$ & $5.92 \times 10^6$ & $7.10 \times 10^6$ & $8.28 \times 10^6$ & $9.47 \times 10^6$ \\
		\hline
	\end{tabular}
\end{table*}

\begin{table*}[bt!]
	\centering
	\renewcommand{\arraystretch}{1.3}
	\caption{Numerical complex multiplications and divisions of the proposed and the competing algorithm against $N_{\rm{t}}$}
	\label{table:complex-number-SNR-Nt}
	\begin{tabular}{|c|c|c|c|c|c|c|}
		\hline
		\multicolumn{2}{|c|}{$N_{\rm{t}}$} & 64 & 128 & 256 & 512 & 1024 \\
		\hline
		\multirow{3}{*}{PE-AltMin ($M=N_{\rm{S}}$)} & $\bar{N}_{iter,\text{F}}^{\text{PE}}$ & 73.49 & 102.80 & 172.11 & 135.96 & 90.81 \\
		\cline{2-7}
		& $\bar{N}_{iter,\text{W}}^{\text{PE}}$ & 64.95 & 64.66 & 64.74 & 64.01 & 64.34 \\
		\cline{2-7}
		& $N_{\rm{Mul}}$ & $2.83 \times 10^6$ & $7.41 \times 10^6$ & $2.40 \times 10^7$ & $3.74 \times 10^7$ & $4.95 \times 10^7$ \\
		\hline
		\multirow{2}{*}{CWAP-AGHC ($M=N_{\rm{S}}$)} & $\bar{N}_{iter,\text{W}}^{\text{AG}}$ & 52.52 & 52.75 & 52.49 & 52.32 & 52.93 \\
		\cline{2-7}
		& $N_{\rm{Mul}}$ & $2.03 \times 10^6$ & $2.91 \times 10^6$ & $6.23 \times 10^6$ & $1.92 \times 10^7$ & $7.02 \times 10^7$ \\
		\hline
		\multirow{3}{*}{PE-AltMin ($M>N_{\rm{S}}$)} & $\bar{N}_{iter,\text{F}}^{\text{PE}}$ & 63.37 & 87.47 & 116.33 & 141.05 & 150.81 \\
		\cline{2-7}
		& $\bar{N}_{iter,\text{W}}^{\text{PE}}$ & 47.13 & 46.61 & 45.64 & 44.08 & 44.22 \\
		\cline{2-7}
		& $N_{\rm{Mul}}$ & $1.18 \times 10^6$ & $3.15 \times 10^6$ & $8.23 \times 10^6$ & $1.98 \times 10^7$ & $4.22 \times 10^7$ \\
		\hline
		\multirow{2}{*}{CWAP-AGHC ($M>N_{\rm{S}}$)} & $\bar{N}_{iter,\text{W}}^{\text{AG}}$ & 29.36 & 29.36 & 29.17 & 29.04 & 29.22 \\
		\cline{2-7}
		& $N_{\rm{Mul}}$ & $1.69 \times 10^6$ & $2.57 \times 10^6$ & $5.89 \times 10^6$ & $1.88 \times 10^7$ & $6.99 \times 10^7$ \\
		\hline
	\end{tabular}
\end{table*}

\begin{table*}[bt!]
	\centering
	\renewcommand{\arraystretch}{1.3}
	\caption{Numerical complex multiplications and divisions of the proposed and the competing algorithm against $N_{\rm{r}}$}
	\label{table:complex-number-SNR-Nr}
	\begin{tabular}{|c|c|c|c|c|c|c|}
		\hline
		\multicolumn{2}{|c|}{$N_{\rm{r}}$} & 1 & 2 & 4 & 8 & 16 \\
		\hline
		\multirow{3}{*}{PE-AltMin ($M=N_{\rm{S}}$)} & $\bar{N}_{iter,\text{F}}^{\text{PE}}$ & 32.73 & 32.29 & 70.85 & 170.72 & 169.73 \\
		\cline{2-7}
		& $\bar{N}_{iter,\text{W}}^{\text{PE}}$ & 8.00 & 8.00 & 19.05 & 51.83 & 65.16 \\
		\cline{2-7}
		& $N_{\rm{Mul}}$ & $4.56 \times 10^6$ & $4.50 \times 10^6$ & $9.87 \times 10^6$ & $2.38 \times 10^7$ & $2.37 \times 10^7$ \\
		\hline
		\multirow{2}{*}{CWAP-AGHC ($M=N_{\rm{S}}$)} & $\bar{N}_{iter,\text{W}}^{\text{AG}}$ & 8.00 & 14.87 & 25.56 & 41.74 & 52.87 \\
		\cline{2-7}
		& $N_{\rm{Mul}}$ & $2.72 \times 10^5$ & $5.48 \times 10^5$ & $1.12 \times 10^6$ & $2.36 \times 10^6$ & $6.24 \times 10^6$ \\
		\hline
		\multirow{3}{*}{PE-AltMin ($M>N_{\rm{S}}$)} & $\bar{N}_{iter,\text{F}}^{\text{PE}}$ & 32.73 & 32.52 & 107.51 & 117.99 & 114.90 \\
		\cline{2-7}
		& $\bar{N}_{iter,\text{W}}^{\text{PE}}$ & 8.00 & 8.00 & 11.64 & 35.40 & 45.50 \\
		\cline{2-7}
		& $N_{\rm{Mul}}$ & $2.31 \times 10^6$ & $2.29 \times 10^6$ & $7.60 \times 10^6$ & $8.34 \times 10^6$ & $8.13 \times 10^6$ \\
		\hline
		\multirow{2}{*}{CWAP-AGHC ($M>N_{\rm{S}}$)} & $\bar{N}_{iter,\text{W}}^{\text{AG}}$ & 8.00 & 14.78 & 16.55 & 24.69 & 29.10 \\
		\cline{2-7}
		& $N_{\rm{Mul}}$ & $2.72 \times 10^5$ & $5.47 \times 10^5$ & $1.10 \times 10^6$ & $2.29 \times 10^6$ & $5.89 \times 10^6$ \\
		\hline
	\end{tabular}
\end{table*}

Fig. \ref{fig:sim-SE-Nt} describes the relationship between the achievable sum-rate and the number of transmitted antennas ${{N}_{\text{BS}}}$ when ${{N}_{\text{U}}}=16$, $U=4$ and SNR = 10 dB. The sum-rate is positively correlated with ${{N}_{\text{BS}}}$. When $M=N_{\rm{S}}=4$, the sum-rate of the proposed CWAP-AGHC scheme and the PE-AltMin algorithm are approximately equal. When $M=4$ and $N_{\rm{S}}=2$, the sum-rate of the proposed CWAP-AGHC scheme is higher than that of the PE-AltMin algorithm. Table \ref{table:complex-number-SNR-Nt} lists the number of complex multiplications and divisions with ${{N}_{\text{BS}}}$. Compared with the PE-AltMin algorithm, complexity of the proposed CWAP-AGHC scheme is reduced when ${{N}_{\text{BS}}}\le 512$.

Fig. \ref{fig:sim-SE-Nr} illustrates the achievable sum-rate with the number of received antennas ${{N}_{\text{U}}}$ when ${{N}_{\text{BS}}}=256$, $U=4$ and SNR = 10 dB. With the increase of ${{N}_{\text{U}}}$, the sum-rate increases gradually. When $M=N_{\rm{S}}$, the sum-rate of the proposed CWAP-AGHC scheme is approximately same as that of the PE-AltMin algorithm. When $M>N_{\rm{S}}$, the sum-rate of the proposed CWAP-AGHC scheme is slightly higher than that of the PE-AltMin algorithm. Table \ref{table:complex-number-SNR-Nr} lists the number of complex multiplications and divisions with ${{N}_{\text{BS}}}$. Complexity of the proposed CWAP-AGHC scheme is much lower than that of the PE-AltMin algorithm.

\section{Conclusion}

To reduce complexity of existing AltMin hybrid precoding algorithms, the CWAP-AGHC scheme was proposed in this paper. Firstly, the expression of the SE was simplified to a function mainly depending on the analog precoder. Then the simplified SE was further converted to the sum of a series of sub-rates. The original SE maximizing problem was transformed into a series of sub-rate maximization problem. Upon maximizing each sub-rate, the closed-form solution of the analog precoder was given directly without iterations. Next, our previously-proposed AG algorithm was adopted to design the hybrid combiner for each user. Simulation results have shown that compared with the AltMin algorithms, the proposed CWAP-AGHC scheme can achieve approximately same or higher sum-rate with lower complexity when the number of RF chains is equal to or larger than the number of data streams.

%\cite{9749240}

%\bibliography{test}

%\begin{IEEEbiographynophoto}{Jane Doe}
%	Biography text here without a photo.
%\end{IEEEbiographynophoto}
%
%\begin{IEEEbiography}[{\includegraphics[width=1in,height=1.25in,clip,keepaspectratio]{fig1.png}}]{IEEE Publications Technology Team}
%	In this paragraph you can place your educational, professional background and research and other interests.\end{IEEEbiography}

\end{document}